\documentclass[12pt]{article}

\usepackage{amsmath,amssymb,cite,bm}
\usepackage{graphicx}

\makeatletter
 
  \@addtoreset{equation}{section}
 \makeatother

\newcommand{\n}{\nonumber \\}
\newcommand{\Tr}{\mathrm{Tr}}
\newcommand{\tr}{\mathrm{tr}}
\newcommand{\diag}{\mathrm{diag}}
\newcommand{\cN}{\mathcal{N}}
\newcommand{\cL}{\mathcal{L}}

\newcommand{\threej}[6]{{\small
			\begin{pmatrix}
			#1 &\hspace{-2mm} #2 &\hspace{-2mm} #3 \\
			#4 &\hspace{-2mm} #5 &\hspace{-2mm} #6
			\end{pmatrix}
			}}
\newcommand{\scriptthreej}[6]{{\scriptsize
			\begin{pmatrix}
			#1 &\hspace{-2mm} #2 &\hspace{-2mm} #3 \\
			#4 &\hspace{-2mm} #5 &\hspace{-2mm} #6
			\end{pmatrix}
			}}
\newcommand{\sixj}[6]{{\small
			\begin{Bmatrix}
			#1 &\hspace{-2mm} #2 &\hspace{-2mm} #3 \\
			#4 &\hspace{-2mm} #5 &\hspace{-2mm} #6
			\end{Bmatrix}
			}}
\newcommand{\scriptsixj}[6]{{\scriptsize
			\begin{Bmatrix}
			#1 &\hspace{-2mm} #2 &\hspace{-2mm} #3 \\
			#4 &\hspace{-2mm} #5 &\hspace{-2mm} #6
			\end{Bmatrix}
			}}

\newcommand{\be}{\begin{equation}}
\newcommand{\ee}{\end{equation}}
\newcommand{\bea}{\begin{eqnarray}}
\newcommand{\eea}{\end{eqnarray}}
\newcommand{\beann}{\begin{eqnarray*}}
\newcommand{\eeann}{\end{eqnarray*}}

\newcommand{\ba}{\begin{array}}
\newcommand{\ea}{\end{array}}

\allowdisplaybreaks

\begin{document}

\setlength{\oddsidemargin}{0cm}
\setlength{\baselineskip}{7mm}

\begin{titlepage}
\renewcommand{\thefootnote}{\fnsymbol{footnote}}
\begin{normalsize}
\begin{flushright}
\begin{tabular}{l}
KUNS-2422
\end{tabular}
\end{flushright}
  \end{normalsize}

~~\\

\vspace*{0cm}
    \begin{Large}
       \begin{center}
         {Exact results for perturbative partition functions of \\ 
         theories with $SU(2|4)$ symmetry}
       \end{center}
    \end{Large}
\vspace{0.7cm}

\begin{center}
Yuhma A{\sc sano}\footnote
            {
e-mail address : 
yuhma@gauge.scphys.kyoto-u.ac.jp }, 
Goro I{\sc shiki}\footnote
            {
e-mail address : 
ishiki@gauge.scphys.kyoto-u.ac.jp}, 
Takashi O{\sc kada}\footnote
            {
e-mail address : 
okada@gauge.scphys.kyoto-u.ac.jp }
    {\sc and}
Shinji S{\sc himasaki}\footnote
           {
e-mail address : 
shinji@gauge.scphys.kyoto-u.ac.jp }\\
      
\vspace{0.7cm}
                    
      {\it Department of Physics, Kyoto University}\\
               {\it Kyoto, 606-8502, Japan}\\
          
\end{center}

\vspace{0.7cm}

\begin{abstract}
\noindent

In this paper, we study the 
theories with $SU(2|4)$ symmetry which consist
of the plane wave matrix model (PWMM), super Yang-Mills
theory (SYM) on $R\times S^2$ and SYM on $R\times S^3/Z_k$.
The last two theories can be realized as theories
around particular vacua in PWMM, 
through the commutative limit of fuzzy sphere
and Taylor's T-duality.
We apply the localization method to PWMM 
to reduce the partition function and 
the expectation values of a class of supersymmetric operators 
to matrix integrals.
By taking the commutative limit and performing the T-duality,
we also obtain the matrix integrals for 
SYM on $R\times S^2$ and SYM on $R\times S^3/Z_k$. 
In this calculation, we ignore possible instanton effects and
our matrix integrals describe the perturbative part exactly.
In terms of the matrix integrals,
we also provide a nonperturbative proof of 
the large-$N$ reduction
for circular Wilson loop operator and free energy in
${\cal N}=4$ SYM on $R\times S^3$.

\end{abstract}
\vfill
\end{titlepage}
\vfil\eject

\setcounter{footnote}{0}

\tableofcontents

\section{Introduction}

Recently, there has been increasing interest in localization in quantum field theory,
which enables us to exactly compute a certain class of physical observables.
The exact computations of 
the partition function and the vacuum expectation value (vev) of a Wilson loop
have been done, for instance, 
in $\cN=2$ or $\cN=4$ supersymmetric Yang-Mills (SYM) theories in four dimensions 
\cite{Nekrasov:2002qd,Pestun:2007rz}
and $\cN=2$ quiver Chern-Simons-matter theories in three dimensions \cite{Kapustin:2009kz}.
These exact results have not only provided a nontrivial evidence of AdS/CFT duality, 
but also revealed a surprising relationship between 
$\cN=2$ SYM on $S^4$ and Liouville/Toda CFT \cite{Alday:2009aq,Wyllard:2009hg}.
More recently,
the localization was also applied to the computation of the partition function 
of $\cN=1$ SYM in five dimensions to examine its relation to M5-brane 
\cite{Kallen:2012cs,Hosomichi:2012ek,Kim:2012av,Kim:2012qf}.

The localization technique should be useful 
also for a matrix quantum mechanics or a matrix model
since they, in general, involve complicated interactions.
For instance, the partition functions of matrix models of Yang-Mills type in zero dimension
having $[X_m,X_n]^2$ interactions were computed by using the localization 
in \cite{Moore:1998et,Kazakov:1998ji}.
In this paper, we apply the localization to the plane wave matrix model (PWMM) 
\cite{Berenstein:2002jq},
which was originally proposed as a matrix quantum mechanics 
describing M-theory on the pp-wave spacetime in the light cone frame.
This theory is a mass deformation of the BFSS matrix theory \cite{BFSS}
with maximal supersymmetries preserved.
In contrast to the BFSS matrix theory, 
PWMM has no flat directions because of the mass deformation.
The vacua of PWMM are discrete and given by fuzzy spheres.

PWMM is also known as one of theories with $SU(2|4)$ symmetry \cite{Lin:2005nh}, 
which consist of
$\mathcal{N}=4$ SYM on $R\times S^3/Z_k$, 2+1 SYM on $R\times S^2$ \cite{Maldacena:2002rb} 
and PWMM.
All these theories are obtained from $\mathcal{N}=4$ SYM on $R\times S^3$
by a consistent truncation and have common features that
they have mass gap, discrete spectrum and many discrete vacua.
Among the $SU(2|4)$ symmetric theories the following relations hold (See Fig.~1)
\cite{Maldacena:2002rb,Ishiki:2006yr}\footnote{
Some extensions of these relations have been discussed in 
\cite{Ishii:2007ex,Ishii:2008tm}};
(a) the theory around each vacuum of 2+1 SYM on $R\times S^2$ is equivalent to
the theory around a certain vacuum of PWMM 
and 
(b) the theory around each vacuum of $\mathcal{N}=4$ SYM on $R\times S^3/Z_k$ is equivalent to
the theory around a certain vacuum of 2+1 SYM on $R\times S^2$ 
with an orbifold condition imposed.

The relation (a) shows that 
the commutative limit of concentric fuzzy spheres with different radii in PWMM 
corresponds to multiple monopoles in 2+1 SYM on $R\times S^2$.
If PWMM and 2+1 SYM on $R\times S^2$ are regarded as theories on D0-branes and D2-branes, 
respectively,
then the relation (a) corresponds to the Myers effect \cite{Myers:1999ps}.
Namely, D0-branes become polarized into fuzzy spheres by a background flux.
The commutative limit of fuzzy spheres realizes a D0-D2 bound state.
The monopole charges in 2+1 SYM on $R\times S^2$ 
are identified with the D0-charges in the D0-D2 bound state.

The relation (b) can be regarded as the Taylor's T-duality in gauge theories on D-branes 
\cite{Taylor:1996ik}.
While it was originally proposed for gauge theories on flat spacetime,
 the relation (b) provides an extension to the case of 
a nontrivial $U(1)$ bundle, 
$S^3/Z_k\rightarrow S^2$. 
The orbifolding condition effectively yields 
the circle along which the T-duality is performed.

These relations were shown directly in the gauge theory side 
in \cite{Maldacena:2002rb,Ishiki:2006yr}.
In \cite{Lin:2005nh}, Lin and Maldacena investigated the gauge/gravity duality 
for theories with $SU(2|4)$ symmetry 
and developed a unified method for providing
the gravity dual for each vacuum of these theories.
In this gravity dual picture, 
it was shown that the relations 
(a) and (b) are also satisfied 
\cite{Ishiki:2006yr}\footnote{See also 
\cite{Klose:2003qc,Fischbacher:2004iu,Agarwal:2010tx}
for the integrability structure 
of the theories with $SU(2|4)$ symmetry and 
\cite{Ling:2006up} for the connection to 
the little string theory.}.

By combining (a) and (b), one obtains the following relation \cite{Ishiki:2006yr};
(c) the theory around each vacuum of 
$\mathcal{N}=4$ SYM on $R\times S^3/Z_k$ is equivalent to
the theory around a certain vacuum of PWMM with an orbifolding condition imposed.

In this paper, 
we obtain exact results of PWMM by using the localization.
In addition, by making use of the relations (a) and (c), 
we obtain exact results of 
2+1 SYM on $R\times S^2$ and $\mathcal{N}=4$ SYM on $R\times S^3/Z_k$ 
from PWMM.

\begin{figure}[t]
\begin{center}
\includegraphics[width=15cm]{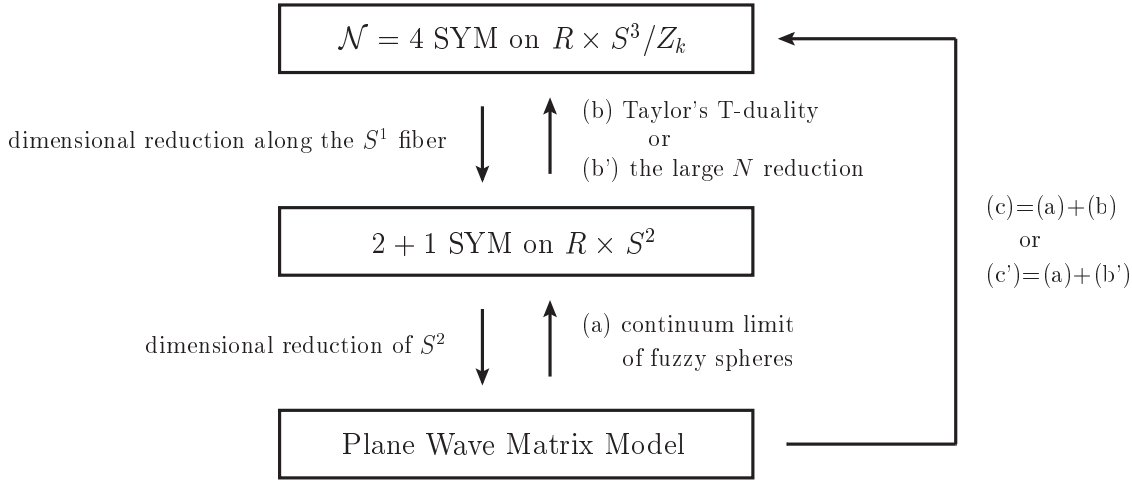}
\caption{The relations among $SU(2|4)$ symmetric theories}
\end{center}
\end{figure}

We perform the localization for PWMM by constructing equivariant cohomology
following \cite{Pestun:2007rz}.
Since PWMM has a noncompact time direction unlike theories considered in \cite{Pestun:2007rz}, 
we have to specify boundary conditions of fields at the future and the past infinities.
In this paper, we demand that all fields are finite at both infinities
such that the action of PWMM is finite.
Once the boundary conditions are specified, the localization can be performed as usual.
We construct off-shell supersymmetries in PWMM,
which is denoted by $Q$ in the following,
and add to the action a $Q$-exact term.
The theory does not depend on the coefficient of the $Q$-exact term.
Sending the coefficient of the $Q$-exact term to infinity
reduces the computation of the vev of $Q$-closed operators
to a one-loop integral around zeros of the $Q$-exact term.
If we ignore the instanton configurations discussed below,
each saddle point is labeled by 
a representation of $SU(2)$ algebra and a constant hermitian matrix $M$.
In the end, the vev of $Q$-closed operators amounts to 
a sum of terms each of which is labeled by an $SU(2)$ representation
and given by a matrix integral of $M$.
Since each vacuum of PWMM is also labeled by an $SU(2)$ representation,
each term in the sum is thought of as the contribution from 
the theory around the corresponding vacuum of PWMM.
We then use the relations (a) and (c) to obtain exact results 
of 2+1 SYM on $R\times S^2$ and $\cN=4$ SYM on $R\times S^3/Z_k$.
By extracting the contribution from the $SU(2)$ representation used in
the relations (a) and (c) from the vev of a $Q$-closed operator in PWMM,
we obtain the vev of the corresponding operators 
in SYM on $R\times S^2$ and SYM on $R\times S^3/Z_k$.

As mentioned above, 
there can be contributions to the saddle points from instantons and anti-instantons
localizing at the past and the future infinities, respectively.
This is reminiscent of the situation in $\cN=2$ or $\cN=4$ SYM on $S^4$ \cite{Pestun:2007rz},
where the instantons and the anti-instantons are localizing at the South and the North poles,
respectively.
In this paper, we simply ignore the instanton contributions.
The evaluation of the instanton part is technically difficult and 
it is beyond the scope of this paper. 
Nevertheless, 
if we restrict ourselves to the 't Hooft limit, where the instantons are suppressed,
our results become exact.

As a consistency check of our computation,
we reproduce a one-loop result of PWMM around the trivial background.
Furthermore, we show that PWMM around the background corresponding to 
$\cN=4$ SYM on $R\times S^3$ through the relation (c)
becomes a Gaussian matrix model.
This is consistent with the results in \cite{Pestun:2007rz,Erickson:2000af,Drukker:2000rr}.

In terms of the matrix integral obtained through the localization in PWMM,
we also check the validity of the nonperturbative formulation of the planar 
$\cN=4$ SYM on $R\times S^3$ proposed in \cite{Ishii:2008ib},
which should be important in the context of
the AdS/CFT correspondence \cite{Maldacena:1997re}.
This formulation is based on the combination of (a) and another relation
(b') in Fig.~1. 
Since the orbifolding condition in (b) needs infinitely large gauge group 
from the beginning,
the relation (c) by itself can not provide a regularization of $\cN=4$ SYM.
However, in the 't Hooft limit,
one has an alternative relation (b') which is based on the 
large-$N$ reduction.
The large-$N$ reduction was first proposed by Eguchi and Kawai 
for theories on flat space\cite{Eguchi:1982nm}
and the relation (b') can be regarded as an extension of 
the large-$N$ reduction 
to the case of a nontrivial $U(1)$ bundle \cite{Ishii:2008ib}.
The large-$N$ equivalence (b') holds only in the planar limit.
However, it does not need the orbifolding condition so that 
its combination with (a), which we call (c') in Fig.~1, 
enables us to regularize the planar
$\cN=4$ SYM on $R\times S^3$ nonperturbatively in terms of PWMM.
It should be remarked that this regularization preserves 16 supersymmetries, 
half of supersymmetries of the original $\cN=4$ SYM.

The validity of the nonperturbative formulation has been checked 
by performing perturbative calculations 
\cite{Ishii:2008ib,Kitazawa:2008mx,Ishiki:2011ct} 
and by numerical simulations \cite{Ishiki:2008te,Ishiki:2009sg}\footnote{
Some preliminary results of numerical simulations of $\cN=4$ SYM in this formulation 
are reported in \cite{Honda:2011qk,Honda:2010nx}, where correlation 
functions and Wilson loops
are numerically computed and compared with the results predicted from the gravity side.}.
In \cite{Ishiki:2011ct}, the vev of the circular Wilson loop in $\cN=4$ SYM 
is also reproduced from PWMM within the ladder approximation.
In this paper, we test this formulation
by computing the free energy and the vev of the 
circular Wilson loop operator nonperturbatively\footnote{
The same kind of the large $N$ equivalence between $\cN=2$ Chern-Simons-matter theories on $S^3$
and their dimensionally reduced models was also investigated 
in \cite{Ishiki:2009vr,Ishiki:2010pe,Asano:2012gt,Honda:2012ni}.}.

This paper is organized as follows.
In Section 2, we review the relations among the theories with $SU(2|4)$ symmetry.
We first perform the consistent truncation of $\cN=4$ SYM on $R\times S^3$
and obtain $\cN=4$ SYM on $R\times S^3/Z_k$, 2+1 SYM on $R\times S^2$ and PWMM.
We then explain how 
$\cN=4$ SYM on $R\times S^3/Z_k$ and 2+1 SYM on $R\times S^2$ can be retrieved from PWMM,
namely, how the relations in Fig.~1 hold.
We also discuss supersymmetric Wilson loops in these relations.
In Section 3, we perform the localization in PWMM.
We construct off-shell supersymmetries in PWMM and add a $Q$-exact term to the action.
After the one-loop integration around saddle points, 
we obtain a matrix integral expression of the partition function 
and the vev of a Wilson loop operator in PWMM.
In Section 4, using the relations (a), (c) and (c'),
we obtain the partition functions and the vev of a Wilson loop operators 
in 2+1 SYM on $R\times S^2$ and $\cN=4$ SYM on $R\times S^3/Z_k$ from those in PWMM.
In Section 5, we summarize our results.
Some useful formulae and perturbative check of our result are summarized in Appendices.

\subsection*{Summary of notations}

The indices used in this paper are summarized as follows;
\begin{align}
M,N,\cdots&=1,2,\cdots, 9,0, \qquad
M',N',\cdots=1,2,\cdots,9, \n
a,b,\cdots&=1,2,3,4, \qquad
a',b',\cdots=2,3,4, \n
m,n,\cdots&=5,6,7,8,9,0, \qquad
m',n',\cdots=5,6,7,8,
\end{align}
where $M,N,\cdots$ are the indices of $SO(9,1)=SO(4)\times SO(5,1)$,
$a,b,\cdots$ are the indices of the $SO(4)$
and $m,n,\cdots$ are the indices of the $SO(5,1)$.

The gauge group of the theories we consider in this paper is always a unitary group.
The ranks of the gauge groups are denoted by $N$, $N_{S^2}$ and $N_{PW}$ 
for $\cN=4$ SYM on $R\times S^3/Z_k$, 2+1 SYM on $R\times S^2$ and PWMM, respectively. The coupling constants for these theories are denoted by 
$g$, $g_{S^2}$ and $g_{PW}$, respectively.

\section{Relations among theories with $SU(2|4)$ symmetries}

In this section, we review the relations among $SU(2|4)$ symmetric theories.
In Section 2.1, we make consistent truncations of $\mathcal{N}=4$ SYM on $R\times S^3$ 
to obtain $\mathcal{N}=4$ SYM on $R\times S^3/Z_k$, $2+1$ SYM on $R\times S^2$ and PWMM,
which all have $SU(2|4)$ symmetry, many discrete vacua and mass gap 
\cite{Kim:2003rza,Lin:2005nh}.
In Section 2.2, we explain 
how higher dimensional theories can be obtained from lower dimensional 
theories 
\cite{Ishiki:2006yr,Ishii:2008ib},
namely, the relations in Fig.~1. 
In Section 2.3, we explain Wilson loops in these relations \cite{Ishiki:2011ct}, 
which are computable by using the localization.

\subsection{Theories with $SU(2|4)$ symmetries}

We start with $\mathcal{N}=4$ SYM on $R\times S^3$.
We follow the same notation as \cite{Pestun:2007rz}.
The metric of $R\times S^3$ and gamma matrices are summarized in Appendix A.
We set the radius of $S^3$ to be $1$.
The action of $\mathcal{N}=4$ SYM on $R\times S^3$ is given by
\begin{align}
S_{R\times S^3}&=\frac{1}{g^2}\int d\tau d\Omega_3 \Tr\Bigl(
-\frac{1}{4}F_{MN}F^{MN}-\frac{1}{2}X_mX^m
-\frac{i}{2}\Psi\Gamma^M D_M \Psi \Bigr), 
\label{SYM on RxS3 1}
\end{align}
where 
\begin{align}
F_{ab}&=\nabla_a X_b-\nabla_b X_a-i[X_a,X_b],\quad
F_{am}=D_aX_m, \quad
F_{mn}=-i[X_m,X_n], \n
D_a&=\nabla_a-i[X_a, \ ], \quad
D_m=-i[X_m, \ ]. 
\label{F in SYM on RxS3}
\end{align}
$a,b,\cdots=1,2,3,4$ are the local Lorentz indices of $SO(4)$
and $m,n,\cdots=5,6,\cdots,9,0$ are the indices of $SO(5,1)$ R-symmetry.
$a=1$ corresponds to $R$ direction, $\tau$, 
while $a=2,3,4$ correspond to $S^3$ direction, $(\theta, \varphi,\psi)$.
$X_a$ are gauge fields, $X_m$ are scalar fields and
$\Psi$ is a Majorana spinor with 16 components.
Because of the conformal coupling to the curvature, 
this theory is massive and the vacuum is trivial and unique.
At the moment, we work in Lorentzian signature, so that $X_0=-X^0$. 
Later, we move to Euclidean signature by regarding $X_0$ as an anti-hermitian matrix\footnote{
As in \cite{Pestun:2007rz}, the integrand of the path integral is defined by
$\exp(S)$. After the Wick rotation, the action $S$ becomes negative definite.}.

For later convenience, we take vielbein as right-invariant 1-form defined in Appendix B
and expand the gauge field on $S^3$ in terms of it.
In this local Lorentz frame, the action takes the form
\begin{align}
S_{R\times S^3}
&=\frac{1}{g^2}\int d\tau d\Omega_3 \Tr\Bigl[
-\frac{1}{2}(\partial_1 X_{b'}-i\cL_{b'}X_1-i[X_1,X_{b'}])^2 \n
&\hspace{35mm}
-\frac{1}{4}(2\varepsilon_{a'b'c'}X_{c'}+i\cL_{a'}X_{b'}-i\cL_{b'}X_{a'}-i[X_{a'},X_{b'}])^2 \n
&\hspace{35mm}
-\frac{1}{2}(D_aX_m)^2-\frac{1}{2}X_mX^m
-\frac{i}{2}\Psi \Gamma^1 \partial_1 \Psi
+\frac{1}{2}\Psi\Gamma^{a'} \cL_{a'} \Psi-\frac{3i}{8}\Psi\Gamma^{234}\Psi \n
&\hspace{35mm}
+\frac{1}{4}[X_m,X_n][X^m,X^n]-\frac{1}{2}\Psi\Gamma^m[X_m,\Psi]
\Bigr],
\label{SYM on RxS3 2}
\end{align}
where $\cL_{a'}$ are the Killing vectors defined in \eqref{Killing vector}.

The action is invariant under the following supersymmetry transformations
\begin{align}
\delta_s X_M&=-i\Psi \Gamma_M \epsilon, \n
\delta_s \Psi&=\left(\frac{1}{2}F_{MN}\Gamma^{MN}
-\frac{1}{2}X_m\tilde{\Gamma}^m\Gamma^a\nabla_a\right)\epsilon.
\label{susy transf}
\end{align}
Here $\epsilon$ is a conformal Killing spinor satisfying 
\begin{align}
\nabla_a \epsilon=\tilde{\Gamma}_a\tilde{\epsilon},
\label{KS eq 1}
\end{align}
where $\tilde{\epsilon}$ is another spinor satisfying
\begin{align}
\Gamma^a\nabla_a \tilde{\epsilon}&=-\frac{1}{2}\epsilon. 
\label{KS eq 2}
\end{align}
Here $\epsilon$ is Grassmann even, so that $\delta_s$ is Grassmann odd.
One can easily solve these equations 
with the ansatz $\tilde{\epsilon}=\pm\frac{1}{2}\Gamma^{19}\epsilon$, for which
\eqref{KS eq 1} and \eqref{KS eq 2} become
\begin{align}
\nabla_a \epsilon=\pm\frac{1}{2}\Gamma^a\Gamma^{19}\epsilon.
\label{KS eq}
\end{align}
Then, the solution is given by
\begin{align}
\epsilon_+=
\begin{pmatrix}
e^{\frac{\tau}{2}} \ \eta_1 \\
e^{\frac{\tau}{2}}  \bar{g} \ \eta_2 \\
e^{-\frac{\tau}{2}} \ \eta_3 \\
e^{-\frac{\tau}{2}}  \bar{g} \ \eta_4
\end{pmatrix} \quad \text{and} \quad
\epsilon_-=
\begin{pmatrix}
e^{-\frac{\tau}{2}} g \ \eta_1 \\
e^{-\frac{\tau}{2}} \ \eta_2 \\
e^{\frac{\tau}{2}} g \ \eta_3 \\
e^{\frac{\tau}{2}} \ \eta_4
\end{pmatrix},
\label{KS}
\end{align}
for the upper and the lower sign in (\ref{KS eq}), respectively.
$\eta_{1,2,3,4}$ are four-component constant spinors and $g$ and $\bar{g}$ are defined by
\begin{align}
g&=e^{\frac{\varphi}{2}J_4}e^{\frac{\theta}{2}J_3}e^{\frac{\psi}{2}J_4}, \n
\bar{g}
&=e^{-\frac{\varphi}{2}\bar{J}_4}e^{-\frac{\theta}{2}\bar{J}_3}e^{-\frac{\psi}{2}\bar{J}_4},
\end{align}
where $J_3$, $\bar{J}_3$, $J_4$ and $\bar{J}_4$ are defined in Appendix A.
For each case, there exist $4\times 4=16$ constant spinors, 
and thus the theory totally possesses 32 supersymmetries. 
Note that, for each case, half of Killing spinors do not depend on the coordinates of $S^3$.
They will, therefore, survive even in theories with $SU(2|4)$ symmetry,
which are obtained by a consistent truncation of $\cN=4$ SYM on $R\times S^3$.
Thus, all the $SU(2|4)$ symmetric theories possess 16 supersymmetries.

\subsubsection*{$\mathcal{N}=4$ SYM on $R\times S^3/Z_k$}

First, we consider $\mathcal{N}=4$ SYM on $R\times S^3/Z_k$. 
The $Z_k$ acts on the $S^1$ fiber of $S^3$.
SYM on $R\times S^3/Z_k$ is therefore obtained by 
making a consistent truncation for the fields
so that only the modes
which have the periodicity $(\theta,\varphi,\psi)\sim (\theta,\varphi,\psi+4\pi/k)$ 
are surviving. 
The action takes the same form as \eqref{SYM on RxS3 1} or \eqref{SYM on RxS3 2}.
The vacuum of this theory is determined by the flat connection on $S^3/Z_k$
and so characterized by the holonomy $U$ along the $S^1$ fiber up to gauge transformation.
Since $\pi_1(S^3/Z_k)=Z_k$, $U$ satisfies $U^k=1$.
Hence $U$ can be written as
\begin{align}
U=\diag(\mathbf{1}_{M_1},
e^{2\pi i/k} \mathbf{1}_{M_2},
e^{2\pi i \times 2/k} \mathbf{1}_{M_3},\cdots,
e^{2\pi i(k-1)/k} \mathbf{1}_{M_k}),
\label{holonomy}
\end{align}
where the sum of the multiplicities is equal to the rank of the gauge group,
$N=\sum_i M_i$.
The vacua of $\mathcal{N}=4$ SYM on $R\times S^3/Z_k$ are parametrized by
a set of the multiplicities $\{M_i| i=1,2,\cdots,k, \; \sum_i M_i=N \}$.

\subsubsection*{2+1 SYM on $R\times S^2$} 

Second, we consider 2+1 SYM on $R\times S^2$.
This theory is easily obtained by 
taking $k\rightarrow \infty$ limit for $\mathcal{N}=4$ SYM on $R\times S^3/Z_k$ 
or just dropping the fiber dependence of the fields in \eqref{SYM on RxS3 2},
\begin{align}
S_{R\times S^2}&=\frac{1}{g_{S^2}^2}\int d\tau d\Omega_2 \Tr\Bigl(
-\frac{1}{2}(\partial_1 X_{b'}-iL^{(0)}_{b'}X_1-i[X_1,X_{b'}])^2 \n
&\hspace{35mm}
-\frac{1}{4}(2\varepsilon_{a'b'c'}X_{c'}+iL^{(0)}_{a'}X_{b'}-iL^{(0)}_{b'}X_{a'}
-i[X_{a'},X_{b'}])^2 \n
&\hspace{35mm}
-\frac{1}{2}(D_aX_m)^2-\frac{1}{2}X_mX^m
-\frac{i}{2}\Psi \Gamma^1 \partial_\tau \Psi
+\frac{1}{2}\Psi\Gamma^{a'} L^{(0)}_{a'} \Psi-\frac{3i}{8}\Psi\Gamma^{234}\Psi \n
&\hspace{35mm}
+\frac{1}{4}[X_m,X_n][X^m,X^n]-\frac{1}{2}\Psi\Gamma^m[X_m,\Psi]
\Bigr), \label{SYM on RxS2 1}
\end{align}
where $L_{a'}^{(0)}$ are ordinary angular momentum operators, 
which are defined in Appendix C. 
It follows from \eqref{metric of S^3} that
the radius of $S^2$ is $\frac{1}{2}$.
One can rewrite $X_{a'}$ in terms of gauge fields and a scalar field on $S^2$
by decomposing $X_{a'}$ into horizontal and vertical components;
\begin{align}
\vec{X}=\Phi \vec{e}_r +a_{2} \vec{e}_\varphi-a_{3}\vec{e}_\theta,
\label{def of Phi}
\end{align}
where $\vec{X}=(X_2,X_3,X_4)$, 
$\vec{e}_r=(\sin\theta\cos\varphi,\sin\theta\sin\varphi,\cos\theta)$, 
$\vec{e}_\theta=(\cos\theta\cos\varphi,\cos\theta\sin\varphi,-\sin\theta)$ and
$\vec{e}_\varphi=(-\sin\varphi,\cos\varphi,0)$.
$a_{2}$ and $a_{3}$ are the gauge fields in the local Lorentz frame
and $\Phi$ is the scalar field on $S^2$.
Then, the first two lines in \eqref{SYM on RxS2 1} are rewritten as
\begin{align}
\int d\tau d\Omega_2\Tr\left(
-\sum_{i=2,3}\frac{1}{2}(f_{1i})^2
-\frac{1}{2}(f_{23}-2\Phi)^2
-\frac{1}{2}(D_{1}\Phi)^2
-\sum_{i=2,3}\frac{1}{2}(D_{i}\Phi)^2
\right),
\end{align}
where $f_{1i} \  (i=2,3)$ and $f_{23}$ are the field strength on $R\times S^2$.
The vacuum of this theory is determined by
\begin{align}
&f_{1i}=0, \quad f_{23}-2\Phi=0, \quad D_{1}\Phi=0, \quad
D_{i}\Phi=0, \quad X_m=0 \qquad (i=2,3).
\end{align}
In the gauge in which $X_1=0$ and $\Phi$ is diagonal, 
the first four equations are solved by
\begin{align}
&\hat{a}_2=0,\quad \hat{a}_3=-\frac{\cos\theta\mp 1}{\sin\theta}\hat{\Phi},\n
&\hat{\Phi}=2 \ \diag(q_{-\Lambda/2}\bm{1}_{N_{-\Lambda/2}},\cdots, 
q_{s} \bm{1}_{N_{s}},\cdots,
q_{\Lambda/2}\bm{1}_{N_{\Lambda/2}}),
\label{vacuum of SYM on RxS2}
\end{align}
where $s=-\Lambda/2,-\Lambda/2+1,\cdots, \Lambda/2$ and $\Lambda$ is an even number.
The sum of the multiplicities is equal to the rank of the gauge group,
$N_{S^2}=\sum_sN_s$.
The upper and lower signs correspond to the patch I ($0\leq \theta < \pi$)
and the patch II ($0<\theta \leq \pi$), respectively.
Each diagonal configuration is nothing but the Dirac monopole with monopole charge $q_s$.
The charge quantization condition imposes $q_s$ to be an integer or 
a half-integer.
One can easily translate the solution \eqref{vacuum of SYM on RxS2} 
into that in terms of $X_{a'}$,
\begin{align}
\hat{X}_2=\frac{1\pm\cos\theta}{\sin\theta}\cos\varphi \cdot \hat{\Phi}, \quad
\hat{X}_3=\frac{1\pm\cos\theta}{\sin\theta}\sin\varphi \cdot \hat{\Phi}, \quad 
\hat{X}_4=\mp \hat{\Phi}. \label{vacuum of SYM on RxS2 2}
\end{align}
These backgrounds are combined with angular momentum operators into
those in a monopole background as
\begin{align}
L_{a'}^{(0)}+\hat{X}_{a'}
=\diag(L_{a'}^{(q_{-\Lambda/2})} \bm{1}_{N_{-\Lambda/2}},\cdots, 
L_{a'}^{(q_s)} \bm{1}_{N_{s}},\cdots,
L_{a'}^{(q_{\Lambda/2})} \bm{1}_{N_{\Lambda/2}}),
\label{vacuum of SYM on RxS2 3}
\end{align}
where $L_{a'}^{(q)}$ is defined in \eqref{monopole angular momentum}.

\subsubsection*{Plane wave matrix model}

Finally, PWMM is obtained by dropping the coordinate dependence of $S^2$
in $2+1$ SYM on $R\times S^2$,
\begin{align}
S_{PW}&=\frac{1}{g_{PW}^2}\int d\tau  \Tr\Bigl(
-\frac{1}{4}F_{MN}F^{MN}
-\frac{1}{2}X_mX^m
-\frac{i}{2}\Psi \Gamma^M D_M \Psi
\Bigr),
\label{action of PWMM}
\end{align}
where
\begin{align}
F_{1M}&=D_1X_M=\partial_1X_M-i[X_1,X_M] \quad (M\neq 1),\n
F_{a'b'}&=2\varepsilon_{a'b'c'}X_{c'}-i[X_{a'},X_{b'}], \quad
F_{a'm}=D_{a'}X_m=-i[X_{a'},X_m], \quad
F_{mn}=-i[X_m,X_n], \n
D_1\Psi&=\partial_1\Psi-i[X_1,\Psi], \quad
D_{a'}\Psi=\frac{1}{4}\varepsilon_{a'b'c'}\Gamma^{b'c'}\Psi-i[X_{a'},\Psi], \quad
D_m\Psi=-i[X_m,\Psi].
\label{F in PWMM}
\end{align}
They are obtained by dropping derivatives in \eqref{F in SYM on RxS3} 
in the right-invariant frame.

When both $X_0$ and $X_1$ are Wick rotated so that the 
theory has the ordinary Lorenzian signature, 
PWMM has $R\times SO(3)\times SO(6)_R$ 
symmetry as the bosonic subgroup
of $SU(2|4)$. The first factor, $R$, corresponds to the translation of the 
$\tau$ direction and the second and the third factors 
corresponds to the rotations for $X_{a'}$ and $X_{m}$, respectively.
In this paper, we will construct an equivariant cohomology with 
respect to the action of a $U(1)$ subgroup of the bosonic subgroup
combined with a gauge transformation.

The vacuum of PWMM is given by the solution to the following equations
\begin{align}
\partial_1 X_{b'}-i[X_1,X_{b'}]&=0, \quad
2\varepsilon_{a'b'c'}X_{c'}-i[X_{a'},X_{b'}]=0, \quad
X_m=0.
\end{align}
In $X_1=0$ gauge, the first two equations are solved by
\begin{align}
X_{a'}=-2L_{a'},
\end{align}
where $L_{a'}$ is a representation of $SU(2)$ algebra; 
$[L_{a'},L_{b'}]=i\varepsilon_{a'b'c'}L_{c'}$.
$L_{a'}$ are in general reducible and can be represented as
\begin{align}
L_{a'}=
 \begin{pmatrix}
  \bm{1}_{N_{-\Lambda/2}}\otimes L_{a'}^{[j_{-\Lambda/2}]} & & & & \\
 & \ddots & & & \\
& & \bm{1}_{N_{s}}\otimes L_{a'}^{[j_{s}]} & & \\
& & & \ddots & \\
& & & & \bm{1}_{N_{\Lambda/2}}\otimes L_{a'}^{[j_{\Lambda/2}]}  
 \end{pmatrix},
\label{vacuum of PWMM}
\end{align}
where $s=-\Lambda/2,-\Lambda/2+1,\cdots, \Lambda/2$
and $\Lambda$ is an even number.
$L^{[j]}_{a'}$ is the spin $j$ representation matrix of $SU(2)$ algebra
and $N_{PW}=\sum_s (2j_s+1)N_s$.

\subsection{$\cN=4$ SYM on $R\times S^3/Z_k$ and 2+1 SYM on $R\times S^2$ from PWMM}

Here we explain the relations in Fig. 1.

\subsubsection{2+1 SYM on $R\times S^2$ from PWMM}
\label{2+1 from PWMM}
First, let us review the relation (a).
In order to see (a), one can utilize the 
harmonic expansion of the two theories in (a).

We first consider the theory expanded around the fuzzy sphere 
background \eqref{vacuum of PWMM}
in PWMM.
To analyze this, it is convenient to decompose the fluctuation fields around the background
into blocks according to the block structure in \eqref{vacuum of PWMM}.
We call the block with size $(N_s\times N_t) \otimes ((2j_s+1)\times(2j_t+1))$ as $(s,t)$-block.
For each block, there is a suitable matrix basis called fuzzy spherical harmonics 
$\hat{Y}_{Jm(jj')}$,
which behave as an irreducible representation of $SU(2)$
under the adjoint action of \eqref{vacuum of PWMM}.
Several properties of fuzzy spherical harmonics are summarized in Appendix D.
For instance, the $(s,t)$-block of scalars $X^{(s,t)}(\tau)$ can be expanded as
\begin{align}
X^{(s,t)}(\tau)
=\sum_{J=|j_s-j_t|}^{j_s+j_t}\sum_{m=-J}^{J}
X_{Jm}^{(s,t)}(\tau)\otimes \hat{Y}_{Jm(j_sj_t)}, \label{mode expansion in PWMM}
\end{align}
where $X_{Jm}^{(s,t)}(\tau)$ is a $N_s\times N_t$ matrix.

Next, let us see the theory expanded around 
the monopole background \eqref{vacuum of SYM on RxS2 2} in 2+1 SYM on $R\times S^2$.
We decompose the fluctuation fields into blocks according to \eqref{vacuum of SYM on RxS2 3}, 
where $(s,t)$-block is now $N_s\times N_t$ matrix. 
Since all the fields are in the adjoint representation, 
the $(s,t)$-block couples with gauge fields 
of the monopole background with monopole charge $q_s-q_t$. 
In this case, a useful basis is the monopole spherical harmonics
defined in Appendix C,
which form a basis of sections of a complex line bundle on $S^2$.
Under the action of the angular momentum operator in the presence of a monopole,
they behave as an irreducible representation of $SU(2)$.
The $(s,t)$-block of scalars $X^{(s,t)}(\tau, \Omega)$ can be expanded as
\begin{align}
X^{(s,t)}(\tau,\Omega)
=\sum_{J=|q_s-q_t|}^{\infty}\sum_{m=-J}^{J}
X_{Jm}^{(s,t)}(\tau) Y_{Jm(q_s-q_t)}(\Omega). \label{mode expansion in SYM on RxS2}
\end{align}
The angular momentum is bounded below 
because the background magnetic field carries nonzero angular momentum.

Notice the similarity between \eqref{mode expansion in PWMM} 
and \eqref{mode expansion in SYM on RxS2}.
The angular momentum of fields in SYM on $R\times S^2$ \eqref{mode expansion in SYM on RxS2}
is bounded below by $|q_s-q_t|$ 
while that in PWMM \eqref{mode expansion in PWMM} is bounded below by $|j_s-j_t|$ 
and also bounded above by $j_s+j_t$.
Thus, in \eqref{mode expansion in PWMM}
we take the limit in which 
\begin{align}
&2j_s+1=n+2q_s  \quad \left(-\frac{\Lambda}{2}\leq s\leq \frac{\Lambda}{2}\right), \quad
n\rightarrow \infty \quad \text{with} \quad 
\frac{g_{PW}^2}{n}=\frac{g_{S^2}^2}{4\pi}=\text{fixed}
\label{S2 from PWMM}
\end{align}
and $N_s$ and $\Lambda$ in PWMM are identified with those in 2+1 SYM on $R\times S^2$.
Under this limit, $|j_s-j_t|=|q_s-q_t|$ and $j_s+j_t\rightarrow \infty$ are realized.
Then one can see that
 \eqref{mode expansion in PWMM} coincides with \eqref{mode expansion in SYM on RxS2}.
This shows that the spectrum of 2+1 SYM on $R\times S^2$ is completely reproduced from PWMM.

It also turns out
that the interaction terms of both theories are coincident in the limit \eqref{S2 from PWMM}.
In the mode expansion in PWMM, 
the coefficients of the interaction terms involve 
the trace of the product of three fuzzy spherical harmonics
\begin{align}
&\hat{\mathcal{C}}_{J_1m_1(j_sj_t)J_2m_2(j_tj_u)J_3m_3(j_uj_s)} \n
&\equiv
\Tr(\hat{Y}_{J_1m_1(j_sj_t)}\hat{Y}_{J_2m_2(j_tj_u)}\hat{Y}_{J_3m_3(j_uj_s)}) \n
&=(-1)^{2j_t+J_1+J_2-J_3} \sqrt{(2J_1+1)(2J_2+1)(2J_3+1)}
\threej{J_1}{J_2}{J_3}{m_1}{m_2}{m_3}
\sixj{J_1}{J_2}{J_3}{j_u}{j_s}{j_t}.
\label{C in PWMM}
\end{align}
Similarly, interaction terms in 2+1 SYM on $R\times S^2$ have
the integral over $S^2$ of the product of three monopole spherical harmonics 
\begin{align}
\mathcal{C}_{J_1m_1q_1J_2m_2q_2J_3m_3q_3}
&\equiv
\int d\Omega Y_{J_1m_1q_1}(\Omega)Y_{J_2m_2q_2}(\Omega)Y_{J_3m_3q_3}(\Omega) \n
&=\sqrt{(2J_1+1)(2J_2+1)(2J_3+1)}
\threej{J_1}{J_2}{J_3}{m_1}{m_2}{m_3}
\threej{J_1}{J_2}{J_3}{q_1}{q_2}{q_3}.
\label{C in SYM on RxS2}
\end{align}
where $\scriptthreej{J_1}{J_2}{J_3}{m_1}{m_2}{m_3}$ 
and $\scriptsixj{J_1}{J_2}{J_3}{j_u}{j_s}{j_t}$ are the Wigner's $3j$- and $6j$-symbol,
respectively.
In the limit \eqref{S2 from PWMM}, by putting $j_u-j_s=q_1$, $j_t-j_u=q_2$ 
and $j_s-j_t=q_3$
and using \cite{vmk}
\begin{align}
\begin{Bmatrix}
a & b & c \\
d+R & e+R & f+R
\end{Bmatrix}
\approx
\frac{(-1)^{a+b+c+2(d+e+f+R)}}{\sqrt{2R}}
\begin{pmatrix}
a & b & c \\
e-f & f-d & d-f
\end{pmatrix},
\end{align}
one can show that
\begin{align}
\sqrt{n}\hat{\mathcal{C}}_{J_1m_1(j_sj_t)J_2m_2(j_tj_u)J_3m_3(j_uj_s)}
\rightarrow
\mathcal{C}_{J_1m_1q_1J_2m_2q_2J_3m_3q_3}.
\label{Chat to C}
\end{align}
By renormalizing the fields in PWMM as, $X\rightarrow \sqrt{n}X$,
one can correctly reproduce all the interaction terms of 2+1 SYM from PWMM.

Thus,  
the theory around \eqref{vacuum of SYM on RxS2 2} of 2+1 SYM on $R\times S^2$ 
is equivalent to
the theory around \eqref{vacuum of PWMM} of PWMM in the limit \eqref{S2 from PWMM}.

\subsubsection{$\cN=4$ SYM on $R\times S^3/Z_k$ from 2+1 SYM on $R\times S^2$}
\subsubsection*{Taylor's T-duality}
Next, let us consider the relation (b) in Fig. 1. It states that
SYM on $R\times S^3/Z_k$ can be equivalently described 
by SYM on $R\times S^2$ around appropriate monopole background with 
the orbifolding condition imposed.
This is an extension of the T-duality in gauge theory a la Taylor to that on 
a $U(1)$ bundle on $S^2$.

Let us first consider SYM on $R\times S^3/Z_k$ around the 
trivial background.
$S^3/Z_k$ can be regarded as an $S^1$-bundle on $S^2$ and
one can make the Kaluza-Klein (KK) expansion along the fiber 
$S^1$ direction.
Since $S^3/Z_k$ is a nontrivial fiber bundle, 
the KK expansion can be made locally.
The theory thus obtained is the theory on $R\times S^2$ 
with infinite number of KK modes.
These KK modes 
are sections of a complex line bundle on $S^2$ and 
can be regarded as 
fluctuations around a monopole background in 2+1 SYM on $R\times S^2$,
where the monopole charge is identified with the KK momentum.
Therefore, $\cN=4$ SYM on $R\times S^3/Z_k$ can be obtained 
by expanding 2+1 SYM on $R\times S^2$ around an 
appropriate monopole background 
so that all the KK modes of $\cN=4$ SYM on $R\times S^3/Z_k$ are reproduced.
This is achieved 
in the following manner.
First, we take the background \eqref{vacuum of SYM on RxS2} in 2+1 SYM on $R\times S^2$ with
\begin{align}
q_s=\frac{ks}{2}, \qquad N_s=N 
\qquad \text{for} \qquad -\infty\leq s \leq \infty,
\label{S3 from S2 bg}
\end{align}
where $\Lambda$ in \eqref{vacuum of SYM on RxS2} is set to infinity from the beginning.
Then, we make the identification among blocks of fluctuations around \eqref{S3 from S2 bg}
as 
\begin{align}
X^{(s,t)}(\tau,\Omega)=X^{(s+1,t+1)}(\tau,\Omega) 
\quad \text{for} \ -\infty< {}^\forall s, {}^\forall t< \infty.
\label{orbifold}
\end{align}
In the end, we can retrieve (an infinite copies of) 
$\cN=4$ $U(N)$ SYM on $R\times S^3/Z_k$ around the trivial background.
In fact, the classical action of SYM on $R\times S^2$ becomes 
equal to that of SYM on $R\times S^3/Z_k$ if the 
infinite multiplicity, $\sum_{s}$, is absorbed by
the renormalization of the coupling
constant as 
\begin{align}
\frac{\pi g^2_{S^2}}{2\sum_s} \rightarrow g^2.
\label{absorption}
\end{align}

The same argument holds for $\cN=4$ SYM on $R\times S^3/Z_k$ 
around a nontrivial background \cite{Ishii:2007ex}.
To realize the theory around the background specified by the holonomy 
(\ref{holonomy}), we introduce a further internal structure to 
$\hat{\Phi}$ in (\ref{vacuum of SYM on RxS2}) by replacing
\begin{align}
q_s {\bf 1}_{N_s} \rightarrow 
{\rm diag}(q_s^{(1)} {\bf 1}_{N^{(1)}_s}, \cdots,
q_s^{(k)} {\bf 1}_{N^{(k)}_s}),
\end{align}
for all $s\in Z$.
Then the appropriate background for (\ref{holonomy}) is given by 
\begin{align}
q_s^{(i)}=\frac{ks}{2}+\frac{(i-1)s}{2}, \;\;\;\; N^{(i)}_s=M_i
\label{tdual for holonomy}
\end{align}
for $i=1,2,\cdots,k$ and $s \in Z$.
By expanding the theory on $R\times S^2$ around 
the monopole background with 
(\ref{tdual for holonomy}), one obtains the theory on 
$R\times S^3/Z_k$ with the holonomy (\ref{holonomy}).

Note that the matrix size of 2+1 SYM has to be infinity 
to perform the T-duality
due to the orbifolding condition \eqref{orbifold}, 
for which $\Lambda\rightarrow \infty$ is necessary.
Thus, this can not be 
applied to 2+1 SYM on $R\times S^2$ 
with finite matrix size.

\subsubsection*{Large-$N$ reduction}
If we restrict ourselves to the planar limit, 
we have an alternative way, the large-$N$ reduction,
to realize the theory on $R\times S^3/Z_k$.
See the relation (b') in Fig. 1.
This method does not need 
the orbifolding condition \eqref{orbifold} and 
hence it can be applied to SYM on $R\times S^2$ 
with finite matrix size. 
This implies that if one finds a good UV regularization 
for SYM on $R\times S^2$, one can also regularize the
planar SYM on $R\times S^3/Z_k$ with the same regularization
by using the large-$N$ equivalence (b').
The matrix size on $R\times S^2$ corresponds to 
the UV cutoff for the momentum along the fiber direction.
As we have seen above, the theory on $R\times S^2$ can be regularized 
by PWMM through the relation (a). Hence, in terms of 
the large-$N$ reduction, the planar SYM on 
$R\times S^3/Z_k$ can be regularized by PWMM as we will see 
in Section \ref{3+1 from PWMM}. 

Let us review the large-$N$ reduction.
It is shown in \cite{Ishii:2008ib} that 
the planar limit of $\cN=4$ SYM on $R\times S^3/Z_k$ around the trivial background 
can be retrieved from 2+1 SYM on $R\times S^2$ in the following way.
We first expand 2+1 SYM around the background \eqref{vacuum of SYM on RxS2} 
with
\begin{align}
q_s=\frac{ks}{2}, \qquad N_s=N 
\qquad \text{for} \qquad -\frac{\Lambda}{2}\leq s \leq \frac{\Lambda}{2}.
\label{S3 from S2 bg 2}
\end{align}
At the end of our calculations, we take the limit in which
\begin{align}
&\Lambda\rightarrow \infty, \quad N\rightarrow \infty, 
\quad \text{with} \quad \frac{\pi g_{S^2}^2N}{2}=g^2N= \text{fixed}.
\label{S3 from S2 limit}
\end{align}
Then, we retrieve the planar limit of 
$\cN=4$ SYM on $R\times S^3/Z_k$ around the trivial background.

The theory around nontrivial background with the holonomy (\ref{holonomy})
would be also obtained by replacing the distribution of 
the monopole charges as in (\ref{tdual for holonomy}).

\subsubsection{$\cN=4$ SYM on $R\times S^3/Z_k$ from PWMM}
\label{3+1 from PWMM}

\subsubsection*{Taylor's T-duality and fuzzy sphere}
It is then clear how one can obtain $\cN=4$ SYM on $R\times S^3/Z_k$ from PWMM.
This is achieved by the relation (c) in Fig. 1 which is 
given by the combination of (a) and (b).
Let us first consider the theory on $R\times S^3/Z_k$ around 
the trivial background.
This theory is realized from PWMM though the relation (c) as follows.
We first expand PWMM around the particular background 
(\ref{vacuum of PWMM}) in which the spin $j_s$ of the $s$-th 
block satisfies $2j_s+1=n+ks$. All the 
multiplicities $N_s$ are set to $N$.
We then impose the orbifolding condition on the fluctuations in PWMM. 
Through (a), the resultant theory is equivalent to 
SYM on $R\times S^2$ with monopole charges (\ref{S3 from S2 bg}) with the 
orbifolding condition (\ref{orbifold}) imposed on the fluctuations.
Then through Taylor's T-duality, this theory is equivalent to 
$U(N)$ $\cN=4$ SYM on $R\times S^3/Z_k$. The coupling constant 
should be renormalized as 
\begin{align}
\frac{2\pi^2g_{PW}^2}{n\sum_s} \rightarrow g^2.
\label{renormalization for gpw and g}
\end{align}

For the theory around the nontrivial background labeled by the holonomy
(\ref{holonomy}), we replace the distribution of the monopole charges 
to (\ref{tdual for holonomy}).

Note that in order to perform the T-duality, 
we have to start with PWMM with infinitely large matrices.
Namely, the formal parameters $n$ and $\Lambda$ should be infinite.

\subsubsection*{Large-$N$ reduction on $S^3/Z_k$}
In order to realize $\cN=4$ SYM on $R\times S^3/Z_k$ from PWMM
with finite matrix size, 
we can make use of the relation (c') in Fig. 1 obtained by combining
the relations (a) and (b');
We expand PWMM around the background \eqref{vacuum of PWMM} with
\begin{align}
2j_s+1=n+ks, \qquad N_s=N 
\qquad \text{for} \qquad -\frac{\Lambda}{2}\leq s \leq \frac{\Lambda}{2} 
\label{SYM on RxS3 from PWMM 1}
\end{align}
and take the limit in which
\begin{align}
&n\rightarrow \infty, 
\quad \Lambda\rightarrow \infty, \quad n-\Lambda\rightarrow \infty, 
\quad N\rightarrow \infty \n
&\text{with} \quad \frac{2\pi^2g_{PW}^2 N}{n}=g^2N=\text{fixed}. 
\label{SYM on RxS3 from PWMM 2}
\end{align}
Then, the planar limit of $\cN=4$ SYM on $R\times S^3/Z_k$ around the trivial background 
is retrieved.
The theory around nontrivial background would be also obtained by 
the modification shown in (\ref{tdual for holonomy}).

Note that before one takes the continuum limit, 
the theory is described by a matrix quantum mechanics 
with finite matrix size. Hence, this relation provides 
a non-perturbative formulation of the planar 
$\cN=4$ SYM on $R\times S^3/Z_k$ in terms of PWMM, 
which is alternative to the lattice formulation.
The parameters $n$ and $\Lambda$ correspond to the UV momentum
cutoffs for the $S^2$ and the $S^1$ directions, respectively.

\subsection{Wilson loop}
Let us consider supersymmetric Wilson loops in $\cN=4$ SYM on $R\times S^3$. 
The supersymmetric Wilson loop in $\cN=4$ SYM on $R\times S^3$ 
takes the form
\begin{align}
W(C)=\frac{1}{N} \Tr P \exp \left(
i\int_0^1 ds \left\{
\dot{x}^\mu(s)e_\mu^a(x(s))X_{a}(x(s)) 
+i|\dot{x}(s)|\Theta^m(s)X_m(x(s))
\right\}\right)
\label{WL}
\end{align}
where the contour $C$ is parametrized by $x^\mu:[0,1]\rightarrow C$
and $\Theta^m(s)$ is a vector satisfying $\eta_{mn}\Theta^m\Theta^n=1$.
In order for the Wilson loop to be invariant under \eqref{susy transf},
a Killing spinor in \eqref{KS} has to satisfy
\begin{align}
\left\{\dot{x}^\mu(s)e_\mu^a\Gamma_a +i|\dot{x}(s)|\Theta^m(s)\Gamma_m\right\}
\epsilon(x)=0.
\label{bps condition for wilson loop}
\end{align}

We parametrize the great circle of $S^3$ by 
\begin{align}
x^\mu(s):(\tau(s),\theta(s),\varphi(s),\psi(s))=(0,0,0,4\pi s).
\label{circular wl}
\end{align}
The Wilson loop on this great circle with $\Theta^m= i \delta^{m0}$ is written as
\begin{align}
W(\text{circle})=\frac{1}{N} \Tr P \exp \left(
2\pi i\int_0^1 ds \left\{
X_{4}(x(s)) -X_0(x(s))
\right\}\right).
\label{circle WL}
\end{align}
Then \eqref{bps condition for wilson loop} becomes
\begin{align}
(\Gamma_4-\Gamma_0)\epsilon=0. 
\label{bps condition for circular wilson loop}
\end{align}
In either case of $\epsilon_+$ or $\epsilon_-$ in \eqref{KS},
$\eta_3=-J_4\eta_1$ and $\eta_4=-\bar{J}_4\eta_2$ solves 
\eqref{bps condition for circular wilson loop}. 
Hence, the Wilson loop along the great circle of $S^3$ is half-BPS.
It is shown in \cite{Pestun:2007rz} that the vev of the circular Wilson loop can be computed
by using the localization as
\begin{align}
\langle W(\text{circle})\rangle
=\frac{1}{N} \langle  \Tr e^{2\pi M}\rangle
\equiv \frac{1}{Z}\int dM \frac{1}{N}\Tr e^{2\pi M} e^{-\frac{4\pi^2}{g^2}\Tr M^2}
\label{GMM}
\end{align}
where $\lambda$ is the 't Hooft coupling.

The Wilson loop in $\cN=4$ SYM on $R\times S^3/Z_k$ takes the same form as \eqref{WL} 
except that the contour is on $R\times S^3/Z_k$ and
only the modes which respect the periodicity 
$(\theta,\varphi,\psi)\sim (\theta,\varphi,\psi+\frac{4\pi}{k})$ are left.
The contour (\ref{circular wl}) is considered as the one on $S^3/Z_k$ 
which winds $k$ times around the nontrivial cycle on $S^3/Z_k$.


One can construct the operators in 2+1 SYM on $R\times S^2$ and PWMM which
are equivalent, through the relations in Fig. 1, to the Wilson loop 
(\ref{WL}) in SYM on $R\times S^3/Z_k$ 
such that its contour is closed in $S^3$ \cite{Ishii:2007sy}.
They are obtained by applying the consistent truncation to 
(\ref{WL}).
For the circular Wilson loop operator \eqref{circle WL},
they can be constructed as follows.
The operator in 2+1 SYM on $R\times S^2$ can be obtained
by dropping the coordinate dependence of the $S^1$-fiber of $S^3$,
\begin{align}
W_{R\times S^2}
&=\frac{1}{N_{S^2}} \Tr \exp \left(2\pi i \left(
X_{4} -X_0
\right)\right)|_{(\tau,\theta,\varphi)=(0,0,0)}. 
\label{circle WL RxS2}
\end{align}
Since we have dimensionally reduced the $S^1$-fiber direction where the Wilson loop was winding,
$X_4$ in \eqref{circle WL RxS2} contains only the vertical component $\Phi$ 
in \eqref{def of Phi}.
Therefore, this operator is just a local operator of scalar fields on $R\times S^2$.
The operator in PWMM is obtained from \eqref{circle WL RxS2} 
by the dimensional reduction and takes the same form,
\begin{align}
W_{PWMM}
=\frac{1}{N_{PW}} \Tr \exp \left(2\pi i\left(
X_{4} -X_0
\right)\right)|_{\tau=0}.
\label{circle WL PWMM}
\end{align}
The operators (\ref{circle WL RxS2}) and
(\ref{circle WL PWMM})
preserve half of the supersymmetries in $SU(2|4)$ symmetric theories.

Note that the matrix sizes $N_{S^2}$ and $N_{PW}$ have to be 
infinite for (b) and (c), so that 
(\ref{circle WL RxS2}) and (\ref{circle WL PWMM}) are not 
well-defined in these cases. However, one can treat the matrix
sizes as formal products,
$N_{S^2}=N\Lambda$ and $N_{PW}=Nn\Lambda $, to 
see the following equivalence \cite{Taylor:1996ik,Ishii:2007sy}.

Based on the relations in Fig. 1, one can show the 
following equivalence 
between the operators \eqref{circle WL},
(\ref{circle WL RxS2}) and
(\ref{circle WL PWMM}).
It follows from the relation (a) that
the vev of \eqref{circle WL RxS2} in 
SYM on $R\times S^2$ around \eqref{vacuum of SYM on RxS2 2}
is equivalent to the vev of \eqref{circle WL PWMM} in PWMM
around \eqref{vacuum of PWMM} in the limit \eqref{S2 from PWMM}.
In addition, it also follows from the relation (c) that
the vev of \eqref{circle WL} in
$\cN=4$ SYM on $R\times S^3/Z_k$ around the vacuum
labeled by (\ref{holonomy})
is equivalent to the vev of \eqref{circle WL PWMM} in
PWMM around the corresponding vacuum with 
the orbifolding condition imposed (See Section 2.2.3).
The similar equivalence holds for the relation (c') 
if one takes the fuzzy sphere vacuum in PWMM with
\eqref{SYM on RxS3 from PWMM 1} and takes 
the continuum limit \eqref{SYM on RxS3 from PWMM 2}.
The statement for (c') was checked explicitly 
in \cite{Ishiki:2011ct} to all orders of 
the perturbation theory within the ladder approximation.

\section{Localization in PWMM}
In this section,  we calculate the partition function of PWMM 
up to the instanton part by applying the localization method. 
We first construct off-shell supersymmetries in PWMM and then add a
SUSY exact term to the action. Then the path integral is dominated by 
the saddle point configuration of the exact term as usual.
The saddle point is given by the fuzzy sphere configuration 
labeled by the representation of $SU(2)$ Lie algebra.
After the one-loop integral is performed, 
the total partition function is given, up to the instanton part,
by a sum over all the representations of $SU(2)$ whose 
dimensions are equal to the matrix size $N_{PW}$ of PWMM,
\begin{align}
Z=\sum_{{\cal R}}Z_{\cal R},
\label{sum of rep}
\end{align}
where ${\cal R}$ is the $N_{PW}$ dimensional 
representation of $SU(2)$ and $Z_{\cal R}$ is
contribution from the corresponding saddle point. 
We will see that, for every ${\cal R}$, $Z_{\cal R}$ is written 
as an eigenvalue integral. 


\subsection{Supersymmetry}
Since PWMM is obtained by the dimensional reduction 
of ${\cal N}=4$ SYM on $R\times S^3$ to one dimension, 
the Killing spinors in PWMM are also obtained from those in
${\cal N}=4$ SYM on $R\times S^3$ by 
dropping the components depending on 
the coordinates of $S^3$.
Namely, $\epsilon_+$ with $\eta_2=\eta_4=0$ and 
$\epsilon_-$ with $\eta_1=\eta_3=0$ in (\ref{KS}) 
are the Killing spinors in PWMM.
We consider only $\epsilon_+$ (with $\eta_2=\eta_4=0$) 
and omit the subscript $+$ in the following.
We impose a further condition on $\epsilon$ that the supersymmetry 
transformation by $\epsilon$
leaves the Wilson loop in PWMM (\ref{circle WL PWMM}) invariant.
Consequently, $\epsilon $ takes the following form
\begin{align}
&\epsilon=
e^{\frac{\tau}{2}\Gamma ^{09}}e^{-\frac{\pi}{4}\Gamma ^{49}}
\begin{pmatrix}
\eta _1\\
0\\
0\\
0
\end{pmatrix},
\label{CKS in PWMM}
\end{align}
where $\eta_1$ is a four-component constant spinor
normalized as $\eta_1 \eta_1=1$.
In fact, the Killing vector 
$v^M=\epsilon \Gamma ^M\epsilon$ constructed from 
(\ref{CKS in PWMM}) has components,
\begin{align}
v^0=2\cosh \tau ,\quad v^4=-2,\quad v^9=2\sinh \tau,
\end{align}
with all the other elements zero.
The field defined by 
\begin{align}
\phi :=v^MX_M
\label{def phi}
\end{align}
is invariant under this supersymmetry
because of (\ref{triality}) and hence 
the Wilson loop (\ref{circle WL PWMM}) defined 
as the exponential of $\phi$ is also invariant.

We extend this supersymmetry to off-shell following \cite{Berkovits:1993hx}.
We introduce seven auxiliary fields $K_i (i=1,2,\cdots,7)$ and 
modify the supersymmetry transformations in PWMM to
\begin{align}
&\delta _sX_M=-i\Psi \Gamma _M\epsilon ,\nonumber \\
&\delta _s\Psi =\frac{1}{2}F_{MN}\Gamma ^{MN}\epsilon 
-X_m\tilde{\Gamma}^{m}\Gamma^{19}\epsilon 
+K^i\nu _i,\nonumber \\
&\delta _sK_i=i\nu _i\Gamma ^MD_M\Psi.
\label{offshellsusy}
\end{align}
Here, $\nu_i$ are bosonic spinors determined by the closure of 
the transformations.
In fact, the closure requires $\nu_i$ to satisfy
\begin{align}
&\epsilon \Gamma ^M\nu _i=0,\nonumber \\
&\frac{1}{2}(\epsilon \Gamma _N\epsilon )\tilde \Gamma ^N_{\alpha \beta}=\nu ^i_\alpha \nu ^i_\beta +\epsilon _\alpha \epsilon _\beta ,\nonumber \\
&\nu _i\Gamma ^M\nu _j=\delta _{ij}\epsilon \Gamma ^M\epsilon .
\label{closedness}
\end{align}
Conversely, if these equations are satisfied,
the supersymmetry (\ref{offshellsusy}) together with 
the bosonic symmetries in PWMM form a closed algebra.
For a given supersymmetry parameter $\epsilon$, the 
spinors $\{\nu_i|i=1,\cdots,7\}$ can
be determined by solving (\ref{closedness}).
When $\epsilon$ is given by (\ref{CKS in PWMM}), the equations 
(\ref{closedness}) are solved by
\begin{align}
&\nu _i=\sqrt{2}
e^{\frac{\tau}{2}\Gamma ^{09}}e^{-\frac{\pi}{4}\Gamma ^{49}}\Gamma ^{i8}
\begin{pmatrix}
\eta _1\\
0\\
0\\
0
\end{pmatrix}. \;\;\;\;\; (i=1,2,\cdots,7)
\label{nu}
\end{align}
This is easily checked by noting that (\ref{CKS in PWMM}) and (\ref{nu})
are equal to constant spinors up to a local Lorentz transformation
represented by $e^{\frac{\tau}{2}\Gamma ^{09}}e^{-\frac{\pi}{4}\Gamma ^{49}}$.
Since these constant spinors satisfy (\ref{closedness}) and the equations
(\ref{closedness}) are Lorentz covariant, 
(\ref{CKS in PWMM}) and (\ref{nu}) also satisfy 
(\ref{closedness}).

In order to make the action of PWMM invariant 
under (\ref{offshellsusy}), quadratic terms,
\begin{align}
\frac{1}{g_{PW}^2}\int d\tau \frac{1}{2}{\rm Tr}K_iK_i,
\end{align}
should be added to the action (\ref{action of PWMM}). 
Since these terms have the wrong sign, $K_i$'s should
be integrated over the imaginary axis.

In the following computation, we put 
\begin{align}
\eta_1=
(1,0,0,0)^T
\end{align}
for simplicity.

For later convenience,
we make a change of variables of the path integral 
for the fermion field.
Since $\{ \Gamma ^{M'}\epsilon,\; \nu ^i|M'=1,\cdots,9,\; i=1,\cdots,7 \}$ 
forms the orthogonal basis of 16 component spinors, 
$\Psi$ can be decomposed as
\begin{align}
\Psi =\Psi _{M'}\Gamma ^{M'}\epsilon +\Upsilon _i\nu ^i.
\label{new basis}
\end{align}
We treat $\{\Psi _{M'},\Upsilon _i\}$ as the new variables
in the path integral.\footnote{
Rigorously speaking, we treat $\Psi_{M'}\sqrt{\epsilon \epsilon }$ and 
$\Upsilon_{i}\sqrt{\epsilon \epsilon }$ as the new variables
to have a trivial Jacobian in the path integral measure.}
The supersymmetry transformations are rewritten using the new variables as
\begin{align}
&\delta _sX_{M'}=-i(\epsilon \epsilon )\Psi _{M'}, \;\;\;
(\epsilon \epsilon )\delta _s\Psi _{M'}=(\delta _\phi +\delta _{U(1)})X_{M'},
\nonumber\\
&(\epsilon \epsilon )\delta _s\Upsilon _i=:H_i, \;\;\;\;\;\;\;\;\;\;\;\;
\delta _sH_i=-i(\epsilon \epsilon )(\delta _\phi +\delta _{U(1)})\Upsilon _{i},
\;\;\;\; \delta _s\phi =0,
\end{align}
where $\phi$ is defined in (\ref{def phi}) and 
$H_i \; (i=1,2,\cdots,7)$ are defined by
\begin{align}
H_i&=(\epsilon \epsilon )K_i+2\nu _i\tilde \epsilon X_0+s_i,\\
s_i&:=\nu _i\left( \frac{1}{2}\sum_{P,Q=1}^9F_{PQ}\Gamma ^{PQ}\epsilon 
-2\sum_{m=5}^9X_{m}\Gamma ^{ m}\tilde \epsilon \right).
\end{align}
$\delta_\phi$ denotes the gauge transformation with parameter $\phi$ and
$\delta_{U(1)}$ is the $U(1)$ transformation,
\begin{align}
&\delta _{U(1)}X_{a'}=-2\varepsilon _{a'b'4}v^{4}X^{b'},\nonumber \\
&\delta _{U(1)}X_{m'}=2(-\delta _{m'}^5X_8+\delta _{m'}^8X_5-\delta _{m'}^7X_6+\delta _{m'}^6X_7),\nonumber \\
&\delta _{U(1)}\Upsilon _i=2(\delta _{i1}\Upsilon _4+\delta _{i2}\Upsilon _3-\delta _{i3}\Upsilon _2-\delta _{i4}\Upsilon _1
+\delta _{i6}\Upsilon _7-\delta _{i7}\Upsilon _6).
\end{align}
This transformation forms a diagonal $U(1)$ subgroup 
of the $SO(3)\times SO(6)_R$ symmetry in PWMM\footnote{In the 
Lorentzian signature that we consider in this paper, 
it is a subgroup of $SO(3)\times SO(5,1)_R$.}.

We also introduce the collective notation,
\begin{align}
X&:=
\begin{pmatrix}
X_{M'}\nonumber \\
(\epsilon \epsilon )\Upsilon _i
\end{pmatrix}, \;\;
X':=
\begin{pmatrix}
-i(\epsilon \epsilon )\Psi _{M'}\\
H_i
\end{pmatrix}.
\end{align}
Then the supersymmetry can be written in a compact form as
\begin{align}
&\delta _sX=X',\;\;\;
\delta _sX'=-i(\delta _\phi +\delta _{U(1)})X, \;\;\;
\delta _s\phi =0.
\end{align}

\subsection{Saddle point}
\label{section qv}

We construct a supersymmetry exact term $\delta _sV$.
The Grassmannian functional $V$ is defined by
\begin{align}
V=\Psi \overline{\delta _s\Psi },
\end{align}
where 
\begin{align}
\overline{\delta _s \Psi}&= \frac{1}{2} F_{MN} \tilde \Gamma^{MN}\epsilon 
+ \frac{1}{2}X_{m}\tilde \Gamma^{am}\nabla_{a} \epsilon - K^{i} \nu_i.
\end{align}
The bar stands for the Hermitian conjugate when
$X_0$ and $K_i$'s are integrated over the imaginary axis and 
are regarded as anti-Hermitian matrices.

The functional $V$ can be expressed 
in terms of $\Psi _{M'}$ and $\Upsilon _i$ defined in (\ref{new basis}) as,
\begin{align}
V
&=\left( D_{M'}(v^Q\bar{X}_Q) +\delta _{U(1)}X_{M'}\right) \Psi ^{M'}
+\bar{H}^{ i}\Upsilon _i,
\end{align}
where $\bar{H}^{i}$ and $\bar X_Q$ are defined as 
the Hermitian conjugates of 
$H_i$ and $X_Q$, respectively. Namely, they are obtained by
flipping signs of $X_0$ and $K_i$ in $H_i$ and $X_Q$.

The bosonic part of $\delta _sV$ is calculated to be
\begin{align}
\delta _sV|_{bos}
&=
-e^{\tau}(D_1X_0+X_0-e^{-\tau}K_5)^2
-e^{-\tau}(D_1X_0-X_0+e^{\tau}K_5)^2
-2c \sum_{a'=2}^4(D_{a'}X_0)^2\nonumber \\ 
&\quad 
-2c \sum_{ i\neq 5}(K^{i})^2
+2c (D_4X_9)^2
+2c [X_0,X_9]^2
+2c \sum_{m'=5}^8 [X_0,X_{m'}]^2
+\mathcal{S}
\nonumber \\
&\quad 
+4\sum _{a=1}^{3}\left[
e^{-\tau }
\left\{ F_{a4}^+-\frac{1}{2}D_a(e^{\tau }X_9)+F_{a+4,8}^+\right\} ^2
+e^{\tau }
\left\{ F_{a4}^-+\frac{1}{2}D_a(e^{-\tau }X_9)-F_{a+4,8}^-\right\} ^2
\right],
\label{pos-def}
\end{align}
where $c$ is just a shorthand notation, $c:=\cosh \tau$.
In the following, we also use $s:=\sinh \tau$. ${\mathcal S}$ is defined by
\begin{align}
\mathcal{S}&=e^{\tau }(X_5+D_1X_5+D_2X_6+D_3X_7+D_4X_8+e^{-\tau }F_{98})^2\nonumber \\
&\quad +e^{-\tau }(X_5-D_1X_5-D_2X_6-D_3X_7+D_4X_8-e^{\tau }F_{98})^2\nonumber \\
&\quad +e^{\tau }(X_6+D_1X_6-D_2X_5+D_3X_8-D_4X_7-e^{-\tau }F_{97})^2\nonumber \\
&\quad +e^{-\tau }(X_6-D_1X_6+D_2X_5-D_3X_8-D_4X_7+e^{\tau }F_{97})^2\nonumber \\
&\quad +e^{\tau }(X_7+D_1X_7-D_2X_8-D_3X_5+D_4X_6+e^{-\tau }F_{96})^2\nonumber \\
&\quad +e^{-\tau }(X_7-D_1X_7+D_2X_8+D_3X_5+D_4X_6-e^{\tau }F_{96})^2\nonumber \\
&\quad +e^{\tau }(X_8+D_1X_8+D_2X_7-D_3X_6-D_4X_5-e^{-\tau }F_{95})^2\nonumber \\
&\quad +e^{-\tau }(X_8-D_1X_8-D_2X_7+D_3X_6-D_4X_5+e^{\tau }F_{95})^2.
\label{pos-def 5}
\end{align}
The covariant derivatives $D_{M}$ in PWMM are defined in (\ref{F in PWMM}).
$F_{ab}^{\pm}$ and $F_{m'n'}^{\pm}$ are selfdual  and anti-selfdual part:
\begin{align}
F_{ab}^{\pm}=\frac{1}{2}(F_{ab}\pm \frac{1}{2}\varepsilon _{abcd}F^{cd}),\quad 
F_{m'n'}^{\pm}=\frac{1}{2}(F_{m'n'}\pm \frac{1}{2}\varepsilon _{m'n'p'q'}
F^{p'q'}).
\end{align}
Here $\varepsilon _{abcd}$ and $\varepsilon _{m'n'p'q'}$ are completely anti-symmetric tensors with 
$\varepsilon _{1234}=1$ and $\varepsilon _{5678}=1$, respectively.
After the Wick rotation, $X_0=iX_0^{(E)}$ and $K_i=iK_i^{(E)}$
for $i=1,2,\cdots,7$,
the bosonic part $\delta _sV|_{bos}$
is given by a sum of positive-definite terms.

Then the saddle point of $\delta_s V$ is determined 
by putting all the terms to be zero.
If we ignore possible instanton configurations discussed below,
the saddle point configuration 
(denoted by putting a hat on the fields, $\hat{X}$) is given,
in the temporal gauge $X_1=0$, by
\begin{align}
&\hat X_0^{(E)}=\frac{M}{c},\quad  
\hat K_5^{(E)}=\frac{M}{c^2}, \quad
\hat{X}_{a'}= -2L_{a'}. \; (a'=2,3,4)
\label{QV=0}
\end{align}
All the other fields are zero at the saddle point. Here $L_{a'} (a'=2,3,4)$ 
are representation matrices of $SU(2)$ generators (\ref{vacuum of PWMM}) and
$M$ is a constant matrix satisfying $[L_{a'},M]=0$ for $a'=2,3,4$. 
It is decomposed as
\begin{align}
M=
 \begin{pmatrix}
  M_{-\Lambda/2}\otimes \bm{1}_{2j_{-\Lambda/2}+1} & & & & \\
 & \ddots & & & \\
& & M_{s}\otimes \bm{1}_{2j_{s}+1} & & \\
& & & \ddots & \\
& & & & M_{\Lambda/2}\otimes \bm{1}_{2j_{\Lambda/2}+1} 
 \end{pmatrix},
\end{align}
where $M_{s} \; (s=-\Lambda/2,-\Lambda/2+1,\cdots,\Lambda/2)$ is an
$N_s\times N_s$ constant
matrix. Thus the saddle point is labeled by the representation of $SU(2)$ 
and $\Lambda+1$ matrices $\{M_s\}$.
One can take a gauge in which all $M_s$ are simultaneously diagonalized. 
We will work in this gauge in the following. 
The eigenvalues of $M_s$ are denoted by $m_{si}$, where $i=1,2,\cdots,N_s$.

The configurations (\ref{QV=0}) are obtained as follows.
The first two in (\ref{QV=0}) can be obtained straightforwardly by 
equating the terms containing $X^{(E)}_0$ and $K_5^{(E)}$ 
in \eqref{pos-def} with zero. Similarly, $K_i=0 \;(i=1,2,3,4,6,7)$ 
and $D_4X_9=0$ follow easily. 
By using the saddle point equations for $F_{ab}$ with $a,b=1,2,3$, 
one  can rewrite the Bianchi identity, $D_{1}F_{23}+D_{2}F_{31}+D_{3}F_{12}=0$,
to
\begin{align}
\sum_{a=1}^{3}D^2_a(cX_9) 
+D_1(F_{67}-F_{58})
+D_2(F_{75}-F_{86})
+D_3(F_{56}-F_{78})=0.
\label{Bianchi}
\end{align}
The sum of the last three terms are calculated as
\begin{align}
&-i[D_1X_6-D_2X_5+D_3X_8,X_7]-i[-D_1X_7+D_2X_8+D_3X_5,X_6]
\nonumber\\
&-i[D_1X_8+D_2X_7-D_3X_6,X_5]-i[-D_1X_5-D_2X_6-D_3X_7,X_8]
\nonumber\\
&= -i\sum_{m=5}^8[cF_{9m},X_m].
\end{align}
The equality follows if one uses the saddle point 
equations coming from ${\mathcal S}$. Thus (\ref{Bianchi}) becomes
\begin{align}
\sum_{a=1}^3D_a^2(c X_9)-\sum_{m'=5}^8[X_{m'},[X_{m'},c X_9]]=0.
\label{cX9}
\end{align}
By multiplying \eqref{cX9} by $c X_9$,
taking the trace and integrating over $\tau$,  one obtains
\begin{align}
\int_{-\infty}^{\infty} d\tau \, \Tr \left[ \sum_{a=1}^3\{ D_a(c X_9)\} ^2-\sum_{m'=5}^8[X_{m'},c X_9]^2\right]=0.
\label{LLX}
\end{align}
The surface term for the partial integration is dropped above.
This is justified as follows. 
The field configurations which diverges at infinity,
$\tau \to \pm\infty$, do not contribute to the path integral 
since the action is infinite with such configurations. 
Hence we impose that all fields are finite
at both infinities, $\tau \to \pm\infty$,
so that the action is finite.
Under this boundary condition, 
$cX_9$ is not necessarily finite at infinity. 
However, if one assumes that $cX_9$ is divergent at infinity, 
there is no solution to the saddle point equations.
Hence one can assume that $c X_9\sim \mathcal{O}(1)$ as $\tau \to \pm \infty$
at the saddle point.
Within this assumption, the surface term is vanishing.

Since all the terms in (\ref{LLX}) are 
positive-definite, each term should vanish. Thus 
in the temporal gauge, one finds that
\begin{align}
X_9=\frac{B_9}{c},\;\;\;
[X_{M},X_9]=0,\;\;\; (M=2,\cdots ,8)
\label{[XM,X9]}
\end{align}
where $B_9$ is a constant matrix.
Similarly, for $m'=5,\cdots ,8$, by combining 
(\ref{[XM,X9]}) and the 
saddle point equations coming from ${\mathcal S}$, one can obtain
\begin{align}
\int d\tau {\rm Tr}\left[
\sum_{a=1}^3(D_aX_{m'})^2-
\sum_{n'=5}^8[X_{n'},X_{m'}]^2
\right]=0,
\end{align}
Applying the same argument as $X_9$ yields 
\begin{align}
X_{m'}=B_{m'}, \;\;\; [X_{M},X_{m'}]=0, \;\;\; (M=2,3,5,6,7,8)
\label{[XM,Xm']}
\end{align}
where $B_{m'}$ are constant matrices.
By substituting \eqref{[XM,X9]} and \eqref{[XM,Xm']} into 
the saddle point equations, (\ref{QV=0}) is obtained.

In addition to (\ref{QV=0}), one should also take into 
account the instanton configurations localizing 
at infinity $\tau \rightarrow \pm \infty$. 
When $\tau $ goes to infinity, some terms 
in (\ref{pos-def}) automatically vanish because 
of the coefficients $e^{\pm \tau}$.
Then the saddle point equations for the remaining terms
in  (\ref{pos-def}) are reduced to (anti-)self-dual equations.
They are solved by the instanton solutions in 
PWMM\cite{Yee:2003ge,Lin:2006tr},
which interpolate different fuzzy sphere 
vacua. Since all fields should take the form of (\ref{QV=0})
for finite $\tau$, 
these instantons should be localized at infinity 
$\tau \rightarrow \pm\infty$.
The evaluation of the contribution from the instanton 
configurations is beyond the scope of this paper and here we 
ignore the instantons.

\subsection{Ghost fields}
In order to make a gauge-fixing, 
we introduce ghost fields,
$(C,C_0,\tilde{C},\tilde{C}_0,b,b_0,a_0,\tilde{a}_0)$,
which obey the following 
BRS transformations,
\begin{align}
&\delta_{B}X=[X,C],\;\;\;\;
\delta_{B}X'=[X',C],
\nonumber\\
&\delta_{B}C=a_0-C^2,\;\;
\delta_{B}\phi=[\phi,C],
\nonumber\\
&\delta_{B}\tilde{C}=b,\;\;\;\;\;\;\;\;\;\;\;\;
\delta_{B}b=[\tilde{C},a_0],
\nonumber\\
&\delta_{B}\tilde{a}_0=i\tilde{C}_0,\;\;\;\;\;\;\;\;
\delta_{B}\tilde{C}_0=-i[\tilde{a}_0,a_0],
\nonumber\\
&\delta_{B}b_0=iC_0,\;\;\;\;\;\;\;\;
\delta_{B}C_0=-i[b_0,a_0], \;\;\;\; \delta_B a_0=0.
\label{BRS transformation}
\end{align}
Our convention is that
$(b,b_0,a_0,\tilde{a}_0)$ are bosonic and
$(C,\tilde{C},C_0,\tilde{C}_0)$ are fermionic. 
The latter anti-commutes with $\Psi$ so that
if $X$ (or $X'$) denotes a fermionic field, the commutator 
in (\ref{BRS transformation}) shall express
the anti-commutator, $-\{X,C\}$ (or $-\{X',C\}$).
The ghost fields with subscript $0$ have only zero modes 
for both $R$ direction and the fuzzy sphere direction 
and they eliminate the zero modes of the ghosts properly 
as we will see shortly.
As in \cite{Pestun:2007rz}, 
$a_0$ should be integrated over the imaginary axis.
The square of $\delta_B$ is a gauge 
transformation with parameter $a_0$,
\begin{align}
\delta_B^2=[\;\; , a_0].
\end{align}

We define the action of the supersymmetry on the ghost fields
as follows,
\begin{align}
\delta_s C = \phi,  \;\;\; \delta_s({\rm the \; other \; ghosts})=0.
\end{align}
Then the combined operator $Q=\delta_s+\delta_B$ 
acts on the fields as,
\begin{align}
&QX=X'+[X,C],\;\;\;\;
QX'=-i(\delta_{\phi}+\delta_{U(1)})X+[X',C],
\nonumber\\
&QC=\phi+a_0-C^2,\;\;\;\;
Q\phi=[\phi,C],
\nonumber\\
&Q\tilde{C}=b,\;\;\;\;\;\;\;\;\;\;\;\;\;\;\;\;\;\;\;\;
Qb=[\tilde{C},a_0],
\nonumber\\
&Q\tilde{a}_0=i\tilde{C}_0,\;\;\;\;\;\;\;\;\;\;\;\;\;\;\;\;
Q\tilde{C}_0=-i[\tilde{a}_0,a_0],
\nonumber\\
&Qb_0=iC_0,\;\;\;\;\;\;\;\;\;\;\;\;\;\;\;\;
QC_0=-i[b_0,a_0], \;\;\;\; Q a_0=0.
\label{Q transformation}
\end{align}
$Q^2$ is given as the sum of the $U(1)$ transformation 
and the gauge transformation with parameter $a_0$,
\begin{align}
Q^2 =R, \;\;\;\; R:=-i \delta_{U(1)}+[\;\;, a_0].
\label{Q square}
\end{align}

The gauge-fixing action and the ghost action are introduced as 
a $Q$-exact form,
\begin{align}
V_{\rm gh}&={\rm Tr} \left[ 
\tilde{C}\left( iF+\frac{\xi_1}{2}b +ib_0  \right)
+C\left( \tilde{a}_0-\frac{\xi_2}{2}a_0  \right)
\right],
\nonumber\\
S_{gh}&=\int d\tau \; QV_{gh}=\int d\tau {\rm Tr} \left[ 
b\left( iF+\frac{\xi_1}{2}b +ib_0  \right)
-\tilde{C}\left( QF+\frac{\xi_1}{2}[\tilde{C},a_0]-C_0 \right)
\right.
\nonumber\\ & \hspace{5cm}
\left.
+(\phi +a_0 -C^2)
\left(\tilde{a}_0 -\frac{\xi_2}{2}a_0 \right) 
-iC \tilde{C}_0
\right],
\label{ghost action}
\end{align}
where $F$ denotes the gauge fixing condition.
In the following computation, we adopt the following 
gauge-fixing condition for the theory expanded around 
the saddle point (\ref{QV=0}).
\begin{align}
F=\hat{D}_a\left(\frac{1}{\cosh \tau} X_a\right).
\end{align}
Here, the background covariant derivative $\hat{D}_a$ is defined by
\begin{align}
\hat{D}_a X :=-i [\hat{X}_a, X ]
\end{align}
for $a=1,2,3,4$, 
where $\hat{X}_1 = i\partial /\partial \tau$ and $\hat{X}_a (a=2,3,4)$
are given by the fuzzy sphere background (\ref{QV=0}).
A similar gauge is taken in \cite{Ishiki:2011ct} 
and it is checked that this condition 
properly eliminates massless modes.

Since the theory does not depend on the parameters $\xi_1$ and $\xi_2$, 
we put $\xi_1=\xi_2=0$ in (\ref{ghost action}). Then, the action is 
reduced to 
\begin{align}
S_{gh}&=\int d\tau {\rm Tr} \left[ 
b\left( iF +ib_0  \right)
+\tilde{C}\hat{D}_a \left( \frac{1}{\cosh \tau}D_a C \right)   
+\tilde{C}C_0
\right.
\nonumber\\ & \hspace{2cm}
\left.
+(\phi +a_0 -C^2) \tilde{a}_0 
-iC \tilde{C}_0
-\tilde{C}\hat{D}_a \left(\frac{1}{\cosh \tau}\Psi \Gamma_a \epsilon \right)
\right].
\end{align}
It is easy to see that the last term does not contribute to any Feynman
diagram, so that one can neglect it.
By integrating $(b_0,\tilde{C}_0,C_0)$, the zero modes
of $(b,C,\tilde{C})$ are eliminated. 
Integration over $b$ produces the gauge fixing constraint $F=0$.
After the Wick rotation of $\phi$ and $a_0$, the integration 
over $\tilde{a}_0$ yields the identification $a_0=-\phi$.
Thus, this action provides an ordinary ghost action 
bilinear in $(\tilde{C},C)$, the gauge 
fixing condition and the identification $a_0=-\phi$.

\subsection{One-loop determinant}
Now we consider the $Q$ transform of $V+V_{gh}$ and perform the 
one-loop 
integration around the saddle point (\ref{QV=0}).
For this purpose, we make a redefinition of the fields as
\begin{align}
\tilde{X}':=X'+[X,C], \;\;\;
\tilde{\phi}&:= \phi +a_0-C^2,
\end{align}
and divide all fields to four groups,
\begin{align}
Z_0&=(X_{M'},\tilde{a}_0, b_0), \;\; 
Z_1=(\Upsilon_i,C,\tilde{C}),
\nonumber\\
Z'_0&=(\tilde{\Psi}_{M'},\tilde{C}_0,C_0), \;\; 
Z'_1=(\tilde{H}_i,\tilde{\phi},b).
\end{align}
These groups form doublets under the $Q$ transformation,
\begin{align}
&QZ_i =Z'_i,  \;\;\; QZ'_i = RZ_i, \;\;\; (i=0,1)
\end{align}
where $R$ is defined in (\ref{Q square}).
We expand the full action given by $S_{PW}-Q(V+V_{gh})$ 
around the saddle point 
configuration (\ref{QV=0}):
$Z_i \rightarrow \hat{Z}_i+Z_i $ and 
$Z'_i \rightarrow \hat{Z'}_i+Z'_i $. 
Then the quadratic term of the fluctuations in $V+V_{gh}$, 
which is needed for the one-loop calculation, is schematically 
written as 
\begin{align}
V^{(2)}=(Z'_0,Z_1)
\left(
\begin{array}{cc}
D_{00} & D_{01} \\
D_{10} & D_{11} \\
\end{array} 
\right)
\left(
\begin{array}{c}
Z_0 \\
Z_1' \\
\end{array} 
\right),
\end{align}
where $D_{ij} (i,j=0,1)$ are certain 
linear differential operators.
Then the quadratic part of the action is
\begin{align}
QV^{(2)}= (RZ_0,Z'_1)
\left(
\begin{array}{cc}
D_{00} & D_{01} \\
D_{10} & D_{11} \\
\end{array} 
\right)
\left(
\begin{array}{c}
Z_0 \\
Z_1' \\
\end{array} 
\right)+
(Z_0',Z_1)
\left(
\begin{array}{cc}
D_{00} & D_{01} \\
D_{10} & D_{11} \\
\end{array} 
\right)
\left(
\begin{array}{c}
Z'_0 \\
RZ_1 \\
\end{array} 
\right).
\label{quadratic action}
\end{align}
Hence, the one-loop integral produces 
\begin{align}
Z_{\rm 1-loop}=\left(\frac{{\rm det}_{V_{Z_1}}R}{{\rm det}_{V_{Z_0}}R} \right)
^{\frac{1}{2}},
\label{detR over detR}
\end{align}
where the determinants are taken in the functional spaces 
of fluctuations of $Z_1$ or $Z_0$, denoted by $V_{Z_1}$ or $V_{Z_0}$, 
respectively.
Recall that we have adopted the boundary condition 
that the fields are finite at infinity so that the 
action is finite. 
This condition implies that 
when the fields $Z_1$ and $Z_0$ are expanded around 
the saddle point (\ref{QV=0}), 
the fluctuations should vanish at infinity since they are massive.
Hence $V_{Z_1}$ and $V_{Z_0}$ are defined
to be the linear span of field configurations
vanishing at infinity.

Note that there exists a natural linear map $D_{10}$ from 
$V_{Z_0}$ to $V_{Z_1}$ which commutes with $R$.
Then the determinants in (\ref{detR over detR}) 
cancel between ${\rm Im}D_{10}\subset V_{Z_1}$ 
and ${\rm Im}D_{10}^* \subset V_{Z_0}$. Here, 
$D_{10}^*$ is the adjoint operator of $D_{10}$ which 
is obtained by the partial integration in 
the action (\ref{quadratic action}).
Hence, $Z_{\rm 1-loop}$ is reduced to 
\begin{align}
Z_{\rm 1-loop}=\left(\frac{{\rm det}_{{\rm coker}D_{10}}R}
{{\rm det}_{{\rm ker}D_{10}}R} \right)
^{\frac{1}{2}}.
\end{align}

Since $R$ and $D_{10}$ commute, 
the kernel and the cokernel can be decomposed
to a direct sum of the eigenspaces of $R$. From (\ref{Q square}) and
the identification $a_0=-\phi=-2iM+4L_4$, we see that 
eigenvalue $r_i$ of $R$
is written as the sum of eigenvalue of
$[-2iM+4L_4, \;\;]$ and $U(1)$ charge. The decomposition is expressed as
\begin{align}
{\rm ker}D_{10}= \bigoplus_{i}V_{r_i}, \;\;
{\rm coker}D_{10}= \bigoplus_{i}V'_{r_i},
\end{align}
where $V_{r_i}$ and $V'_{r_i}$ are the restrictions of 
the kernel and the cokernel, respectively, to the eigenspace 
of $R$ with eigenvalue $r_i$. 
The one-loop determinant is then written as
\begin{align}
Z_{\rm 1-loop}= 
\prod_{i}r_i^{({\rm dim}V'_{r_i}-{\rm dim}V_{r_i})/2}.
\label{z pert}
\end{align}
Thus, computing $Z_{\rm 1-loop}$ amounts to
finding the index of $D_{10}$ for each eigenspace of $R$.
Note that since our model is one dimensional matrix model, 
both ${\rm ker}D_{10}$ and ${\rm coker}D_{10}$ are finite dimensional.
Hence $D_{10}$ is Fredholm and the index is well-defined.

In the following, we compute the dimensions of these spaces 
for the theory expanded around the saddle point (\ref{QV=0}).
For this purpose, we first describe how to compute it for 
a general class of linear differential operators on the real 
line\footnote{It will not cause any confusion to 
use the letter $R$ both for the real line and $Q^2$ in 
(\ref{Q square}). } $R$.
We consider the set of all $n$ dimensional 
vector valued smooth functions on $R$ which vanish at infinity, 
$S:=\{f: R \rightarrow  C^n | 
\lim_{\tau \rightarrow \pm \infty}f(\tau)= 0 \}$.
We introduce a linear differential operator $D$ acting on the 
vector space $S$ as
\begin{align}
Df(\tau):=\frac{\partial f}{\partial \tau}(\tau) +(A\cdot f)(\tau),
\label{df=0}
\end{align}
where $f \in S$ and $A$ is a smooth function from $R$ to 
the space of $n\times n$ complex matrices.
The product between $f$ and $A$ is defined as usual,
$(A\cdot f)_i(\tau):=A_{ij}(\tau)f_j(\tau)$.
We assume that $A$ has definite limit values,
$\lim_{\tau\rightarrow \pm \infty} A_{ij}(\tau)<\infty$ 
for any $i,j=1,\cdots,n$,
and $A(\tau)$ can be diagonalized for any $\tau\in  R$ as 
\begin{align}
V^{-1}(\tau)A(\tau)V(\tau)=A_d(\tau)
:={\rm diag}(\lambda_1(\tau), \cdots, \lambda_n(\tau))
\label{diagonalize}
\end{align}
by a certain $V \in \Gamma(E)$, where 
$E$ is $SL(n,C)$ bundle on $R$ and $\Gamma(E)$ is the set of 
all smooth sections of $E$.
Since $A(\tau)$ is constant at infinity, 
$\lim_{\tau\rightarrow \pm \infty}\lambda_i(\tau)$ are also 
constants for $1\leq i \leq n$. Then 
$\lim_{\tau\rightarrow \pm \infty}V(\tau)< \infty$ are also constant matrices.
Suppose that $k$ $(1\leq k \leq n)$ eigenvalues in (\ref{diagonalize})
satisfy both 
\begin{align}
\lim_{\tau \rightarrow \infty} {\rm Re}\lambda_i(\tau)>0 \;\;\; {\rm and} 
\;\;\;
\lim_{\tau \rightarrow -\infty} {\rm Re}\lambda_i(\tau)<0 
\label{hantei}
\end{align}
and the other $n-k$ do not.
Then, the dimension of kernel of $D$ is given by the formula,
\begin{align}
{\rm dim}({\rm ker} D) = k.
\label{dimker}
\end{align}

One can show (\ref{dimker}) as follows. 
Since $D$ is the gauge covariant derivative on $R$, 
the gauge field $A$ can be transformed to any value by 
inhomogeneous gauge transformation. Consider such a transformation
by $U\in \Gamma(E)$ 
which maps $A$ to the right-hand side of (\ref{diagonalize}),
\begin{align}
U^{-1}AU+U^{-1}\partial U =A_d.
\end{align}
Such $U$ can be formally written using a path ordering product.
Then, the differential equation $Df=0$ is solved by
\begin{align}
f(\tau)=U(\tau)\exp\left(-\int^{\tau}_0 A_d(\tau')d\tau' \right)f_0,
\label{formal solution}
\end{align}
where $f_0$ is a constant vector.
(\ref{formal solution}) has to vanish at infinity,
for $f \in S$. Since $U$ should
converge to $V$ at infinity, 
$U$ goes to a constant matrix.
By multiplying the inverse of $U$ at infinity, 
the vanishing condition of (\ref{formal solution}) 
at infinity implies (\ref{dimker}).
Indeed, when $k$ of $\lambda_i$'s satisfy 
(\ref{hantei}),
$k$ components of $f_0$ can be nonzero keeping $f(\tau)$ 
vanishing at infinity. This means that the 
dimension of ${\rm ker}D$ is equal to $k$.

Then let us apply the above argument to 
the plane wave matrix model.
The relevant part of the action, $Z_1D_{10}Z_0$, is 
\begin{align}
&2s_i \Upsilon_i+i\tilde{C}(F+b_0)+C\tilde{a}_0
\nonumber\\
&-\frac{i}{\epsilon \epsilon}
\left( 
\delta_{U(1)}X_{M'}
-2i[\hat{X}_{M'},v^4X_4+v^9X_9]
-i[X_{M'},-2iM+v^4\hat{X}_4]
\right)
[\hat{X}_{M'},C].
\label{linearized action}
\end{align} 
Since the fields in the hypermultiplet, 
$(X_{m'},\Upsilon_i)$ $ (m'=5,8,7,8, i=1,2,3,4)$, 
decouple from the fields in the vector multiplet in 
(\ref{linearized action}), 
the index is decomposed to a sum of contributions from 
these two sectors.

\subsubsection*{Hypermultiplet}
We first consider the hypermultiplet sector. 
We define complex scalar fields as
\begin{align}
W_1=X_5+iX_8, \;\;\; W_2=X_6+iX_7.
\end{align}
One can read off the action of $D_{10}$ on these fields from 
(\ref{linearized action}).
If $W_1,W_2 \in {\rm ker}D_{10}$, they satisfy
\begin{align}
&\partial W_1 +2i [L_-, W_2]+\frac{s}{c}(W_1+2[L_4,W_1])=0, \nonumber\\
&\partial W_2 -2i [L_+, W_1]+\frac{s}{c}(W_2-2[L_4,W_2])=0,
\label{ker hyper}
\end{align}
where $s=\sinh \tau$ and $c=\cosh \tau$.
We first decompose $W_i (i=1,2)$ to block components 
$\{W_i^{(s,t)} |s,t=-\Lambda/2,\cdots,\Lambda/2 \} $
and then expand each block in terms of the fuzzy spherical harmonics as
\begin{align}
W_i^{(s,t)}=\sum_{J=|j_s-j_t|}^{j_s+j_t}\sum_{m=-J}^{J}
W_{iJm}^{(s,t)}\otimes \hat{Y}_{Jm(j_s,j_t)}.  \;\;\; (i=1,2)
\label{harmonic expansion for W}
\end{align}
By substituting this expansion to (\ref{ker hyper}), we obtain
\begin{align}
&\partial W_{1Jm}^{(s,t)}+\frac{s}{c}(1+2m)W_{1Jm}^{(s,t)}
+2i\delta_- W_{2Jm+1}^{(s,t)}=0, \nonumber\\
&\partial W_{2Jm}^{(s,t)}+\frac{s}{c}(1-2m)W_{2Jm}^{(s,t)}
-2i\delta_+ W_{1Jm-1}^{(s,t)}=0,
\end{align}
where we have defined $\delta_{\pm}= \sqrt{(J\pm m)(J\mp m+1)}$.
It is easy to check that (\ref{hantei}) 
is satisfied only by $W_{1JJ}^{(s,t)}$ and $W_{2J-J}^{s,t}$. 
Indeed, these modes have eigenvalues $(2J+1) \tanh \tau$ which 
satisfy (\ref{hantei}).
The equations for the other modes can be rewritten in the form of 
(\ref{df=0}), where $f=(W_{1Jm}^{(s,t)},W_{2Jm+1}^{(s,t)})^T$ and 
\begin{align}
A=\left(
\begin{array}{cc}
\frac{s}{c}(2m+1) & 2i\delta_-   \\
-2i\delta_- & -\frac{s}{c}(2m+1)    \\
\end{array}
\right)
\label{A hyper}
\end{align}
for $m=-J,-J+1,\cdots,J-1$.
The eigenvalues of (\ref{A hyper}) do not satisfy (\ref{hantei}).
Thus, we find that only $W_{1JJ}^{(s,t)}$ and $W_{2J-J}^{(s,t)}$ 
and their complex conjugates contribute to the index. 

Then, we consider the contribution from $\{\Upsilon_i, i=1,2,3,4\}$ to 
the index. Introducing complex fields, 
\begin{align}
\xi_{1}=\Upsilon_1+i\Upsilon_4, \;\;\; \xi_2=\Upsilon_3+i\Upsilon_2,
\end{align}
and expanding their block components in terms of the spherical harmonics 
as in (\ref{harmonic expansion for W}), one obtains 
\begin{align}
&\partial \xi_{1Jm}^{(s,t)}+\frac{2sm}{c}\xi_{1Jm}^{(s,t)}+2\delta_+
\xi_{2Jm-1}^{(s,t)}=0, \nonumber\\
&\partial \xi_{2Jm}^{(s,t)}-\frac{2sm}{c}\xi_{2Jm}^{(s,t)}+2\delta_-
\xi_{1Jm+1}^{(s,t)}=0,
\end{align}
for $\xi_{1},\xi_2 \in {\rm coker}D_{10} $. Since in this case, there is no 
eigenvalue satisfying (\ref{hantei}), one finds that 
$\{\Upsilon_i, i=1,2,3,4\}$ do not contribute to the index.

In summary, only $W_{1JJ}^{(s,t)}$ and $W_{2J-J}^{(s,t)}$ 
and their complex conjugates contribute to the index for 
the hypermultiplet. The eigenvalues of $R$ for these 
fields are read off from
\begin{align}
RW^{(s,t)}_{1}&=2W_{1}^{(s,t)}+[\hat{\phi},W_{1} ]^{(s,t)}
\nonumber\\
&= \sum_{J,m}2\left\{ (1+2m)W^{(s,t)}_{1Jm}+i(M_sW^{(s,t)}_{1Jm}-
W^{(s,t)}_{1Jm}M_t) \right\}
\otimes \hat{Y}_{Jm(j_s,j_t)}
\end{align}
and so on. Thus, we find that the contribution to (\ref{z pert}) 
from the hypermultiplet is given, up to an overall constant, by 
\begin{align}
\prod_{s,t=-\Lambda/2}^{\Lambda/2}\prod_{J=|j_s-j_t|}^{j_s+j_t}
\prod_{i=1}^{N_s}\prod_{j=1}^{N_t}\frac{1}{(2J+1)^2+(m_{si}-m_{tj})^2}.
\label{z hyper}
\end{align}

\subsubsection*{Vector multiplet}
We then compute contribution from the vector multiplet.
We first calculate the dimension of ${\rm ker}D_{10}$.
If the fields $\{X_M,\tilde{a}_0,b_0 | M=1,2,3,4,9\}$ 
are in ${\rm ker}D_{10}$, they satisfy 
\begin{align}
&F+b_0=0, \label{g1}  \\
&\tilde{a}_0+2\left[\hat{X}_{M'},\frac{1}{\epsilon\epsilon}
[\hat{X}_{M'},v^4X_4+v^9X_9   ]\right]
+\left[
\left[
\hat{X}_{M'},\frac{1}{\epsilon \epsilon}X_{M'}
\right], -2iM+v^4\hat{X}_4 \right]=0, \label{g2} \\
&c(2X_4-i[\hat{X}_2,X_3]+i[\hat{X}_3,X_2])-s(\partial X_4 +i[\hat{X}_4,X_1])
-\partial X_9 =0, \label{g3} \\
& c(\partial X_3 +i[\hat{X}_3,X_1])-s(2X_3+i[\hat{X}_2,X_4]-i[\hat{X}_4,X_2])
-i[\hat{X}_2,X_9]=0, \label{g4} \\
& c(\partial X_2 +i[\hat{X}_2,X_1])-s(2X_2-i[\hat{X}_3,X_4]+i[\hat{X}_4,X_3])
+i[\hat{X}_3,X_9]=0. \label{g5}
\end{align}
Here we reduce the number of equations by partially solving the equations
before applying the argument of eigenvalues.
First, by taking the limit $\tau \rightarrow \pm \infty$ in (\ref{g1}), 
since $F \rightarrow 0$, one obtains $b_0=0$. Then, 
\begin{align}
F=\left[\hat{X}_a, \frac{1}{\cosh \tau}X_a \right]=0,
\label{F=0}
\end{align}
for arbitrary $\tau \in  R$ follows again from (\ref{g1}).
Similarly, $\tilde{a}_0=0$ follows from (\ref{g2}).  
By substituting (\ref{F=0}) to (\ref{g2}), one obtains,
\begin{align}
-\partial \left(
\frac{1}{c}\partial (X_4-sX_9) \right)
+\frac{4}{c}[L_{a'},[L_{a'},X_4-sX_9]]=0.
\label{X4-sX9}
\end{align}
From (\ref{X4-sX9}) we show $X_4-sX_9=0$ as follows.
(\ref{X4-sX9}) takes the form, 
$\partial^2 f -\frac{s}{c}\partial f-4J(J+1)f=0$, where 
$f$ corresponds to $X_4-sX_9$ and $J(J+1)$ is the eigenvalue 
of $[L_{a'},[L_{a'}, \;\; ]]$. Since $f/c$ should vanish at 
infinity, 
\begin{align}
0=\int d\tau \partial \left( \frac{1}{c^2}f\partial f \right)
=\int dx 
\left[ 
\left( \frac{\partial f }{c} \right)^2
+\left(\frac{4J(J+1)-1}{c^2}+\frac{3}{2c^4} \right)f^2
\right].
\end{align}
The right-hand side is a sum of positive definite terms except when $J=0$.
Hence $f=0$ when $J \neq 0$. When $J=0$, the original equation 
is $\partial ((\partial f)/c)=0$ and integrating it under the boundary 
condition, $f/c \rightarrow 0$, yields $f={\rm constant}$.
On the other hand, when $J=0$, the commutator terms in (\ref{g3}) vanish.
By solving this equation together with the conditions, $f=X_4-sX_9={\rm constant}$ and $X_4,X_9 \rightarrow 0 $ $(\tau \rightarrow \pm \infty)$, 
one obtains $X_4=X_9=0$ for $J=0$. In summary, one can put $X_4-sX_9=0$ for
any $J$.

We introduce the complex scalar fields $X_{\pm}=X_2 \pm i X_3$ and 
eliminate $X_4$ by $X_4=sX_9$. Then, the equations (\ref{g1}), 
(\ref{g3}), (\ref{g4}), (\ref{g5}) are written as
\begin{align}
&-i\partial X_1 +i \frac{s}{c}X_1+[L_+,X_-]+[L_-,X_+]+2s[L_4,X_9]=0, 
\nonumber\\
&-[L_+,X_-]+[L_-,X_+]+sX_9-c\partial X_9+2i\frac{s}{c}[L_4,X_1]=0, \nonumber\\
&c(\partial X_+ -2i[L_+,X_1])-s(2X_+ -2[L_4,X_+])-2c^2[L_+,X_9]=0, \nonumber\\
&c(\partial X_- -2i[L_-,X_1])-s(2X_- +2[L_4,X_-])+2c^2[L_-,X_9]=0.
\label{g1345}
\end{align}
We make a redefinition for $X_9$ as $X_9'=c X_9$. 
Note that $X_9'$ does not necessarily vanish at infinity but 
there is no solution to (\ref{g1345}) such that $X_9'$ 
is non-zero at infinity. So one can assume that 
$X_9' \rightarrow 0 $ $(\tau \rightarrow \pm \infty)$. 
We then decompose $X_{\pm}, X_1, X_9'$ into the block components and 
expand them by the fuzzy spherical harmonics as we have done in 
(\ref{harmonic expansion for W}).
For $f=(X^{+(s,t)}_{Jm+1}/\sqrt{2},
X^{-(s,t)}_{Jm-1}/\sqrt{2},iX^{1(s,t)}_{Jm},X^{'9(s,t)}_{Jm})^T$
$(m=-J+1,-J+2,\cdots,J-1, \ J\geq 1)$,
the equations (\ref{g1345}) are written in the form of 
(\ref{df=0}), where $A$ is given by
\begin{align}
A=
\left(
\begin{array}{cccc}
\frac{2ms}{c} & 0 & -\sqrt{2}\delta_- & -\sqrt{2}\delta_- \\
0 & -\frac{2ms}{c} & -\sqrt{2}\delta_+ & \sqrt{2}\delta_+ \\
-\sqrt{2}\delta_- & -\sqrt{2}\delta_+ & -\frac{s}{c} &-\frac{2ms}{c} \\
-\sqrt{2}\delta_- & \sqrt{2}\delta_+ & -\frac{2ms}{c} & -\frac{2s}{c} \\
\end{array}
\right).
\end{align} 
Since $A$ is a real symmetric matrix, the eigenvalues of $A$ are real.
By examining the determinant of $A$ for given $J$ and $m$,
it turns out that $A$ does not have zero eigenvalues for any $\tau$.
Therefore, the sign of each eigenvalue does not change as a function of $\tau$
and we conclude that there is no eigenvalue satisfying (\ref{hantei}).
When $m=J$ or $m=-J$, the equations (\ref{g1345}) are closed
with three fields 
$f=(X^{-(s,t)}_{JJ-1}/\sqrt{2},iX^{1(s,t)}_{JJ},X^{'9(s,t)}_{JJ})^T$
or $f=(X^{+(s,t)}_{J-J+1}/\sqrt{2},iX^{1(s,t)}_{J-J},X^{'9(s,t)}_{J-J})^T$.  
Similarly, we find that the eigenvalues of the $3 \times 3$ 
matrices  do not satisfy (\ref{hantei}).
Hence, the bosonic fields in the vector multiplet do not contribute to the index.

Then let us consider the cokernel of $D_{10}$. 
The elements of ${\rm coker}D_{10}$ in the vector multiplet, 
$(C,\tilde{C},\Upsilon_5,\Upsilon_6,\Upsilon_7)$,
satisfy the following conditions,
\begin{align}
&-\frac{1}{c}\partial \tilde{C}+\frac{1}{c}[iM-2L_4,\partial C]
-8 s [L_4, \Upsilon_5]-8c[L_3,\Upsilon_6]+8c[L_2,\Upsilon_7]=0,  
\nonumber\\
&\frac{1}{c}[L_2,\tilde{C}]-\frac{1}{c}[L_2,[iM-2L_4,C]]
+4ic[L_3,\Upsilon_5]-4is[L_4,\Upsilon_6]-2c\partial \Upsilon_7 -6s\Upsilon_7 =0,\nonumber\\
&\frac{1}{c}[L_3,\tilde{C}]-\frac{1}{c}[L_3,[iM-2L_4,C]]
-4ic[L_2,\Upsilon_5]-4is[L_4,\Upsilon_7]+2c\partial \Upsilon_6 +6s\Upsilon_6 =0,\nonumber\\
&\frac{1}{c}[L_4,\tilde{C}]+\partial \left(\frac{1}{c}\partial C \right) 
-\frac{4}{c}[L_{a'},[L_{a'},C]]-\frac{1}{c}[L_4,[iM-2L_4,C]] 
+2s \partial \Upsilon_5 +6c \Upsilon_5 \nonumber\\
& +4is[L_2,\Upsilon_6]+4is[L_3,\Upsilon_7] =0, 
\nonumber\\
&-s \partial \left(\frac{1}{c}\partial C \right) 
+\frac{4s}{c}[L_{a'},[L_{a'},C]]+2\partial \Upsilon_5
+4i[L_2,\Upsilon_6]+4i[L_3,\Upsilon_7]=0.
\label{coker gauge}
\end{align}
From the coefficients of $\tilde{a}_0$ and $b_0$
in (\ref{linearized action}), we also have 
\begin{align}
\int_{-\infty}^{\infty}d\tau C^{(s,s)}_{00}(\tau)=
\int_{-\infty}^{\infty}d\tau \tilde{C}^{(s,s)}_{00}(\tau)=0,
\label{0 mode is 0}
\end{align}
where the subscript $00$ indicates the zero mode of the fuzzy sphere which 
exists only in the diagonal blocks.
We redefine the fields as 
$\tilde{C}'=(\tilde{C}-[iM-2L_4,C])/(2\sqrt{2} c)$,
$C'=C/c$, $\Upsilon_5'=\sqrt{2}\Upsilon_5$
and also introduce complex fields, 
$\Upsilon_{\pm}=\Upsilon_6 \pm i\Upsilon_7$.
Note that these fields also vanish at infinity.
We also introduce a new field $d= \partial C'$ in order to 
make the equations first order. 
With these variables, (\ref{coker gauge}) is rewritten as
\begin{align}
&\partial C' -d =0,
\nonumber\\
&\partial d +\frac{3s}{c}d +2C'-4[L_{a'},[L_{a'},C']]
+\frac{2\sqrt{2}}{c^2}[L_4,\tilde{C}'] 
+\frac{3\sqrt{2}}{c^2}\Upsilon'_5 =0,
\nonumber\\
&\partial \Upsilon_+ -\sqrt{2}i[L_+,\tilde{C}']
-\sqrt{2}i[L_+,\Upsilon_5']+\frac{3s}{c}\Upsilon_+ 
-\frac{2s}{c}[L_4,\Upsilon_+]=0,
\nonumber\\
&\partial \Upsilon_- +\sqrt{2}i[L_-,\tilde{C}']
-\sqrt{2}i[L_-,\Upsilon_5']+\frac{3s}{c}\Upsilon_- 
+\frac{2s}{c}[L_4,\Upsilon_-]=0,
\nonumber\\
&\partial \tilde{C}' +\frac{2s}{c}\tilde{C}' 
+\frac{2s}{c}[L_4,\Upsilon_5']-\sqrt{2}i([L_+,\Upsilon_-]-[L_-,\Upsilon_+])=0,
\nonumber\\
&\partial \Upsilon_5' +\frac{2s}{c}[L_4,\tilde{C}']
+\frac{3s}{c}\Upsilon_5'
+\sqrt{2}i([L_+,\Upsilon_-]+[L_-,\Upsilon_+])=0.
\end{align}
We decompose the fields to the block components and 
expand them by the harmonics. 
Then $f=(C'{}^{(s,t)}_{Jm},d^{(s,t)}_{Jm},
\Upsilon^{+(s,t)}_{Jm+1},
\Upsilon^{-(s,t)}_{Jm-1},
\Upsilon'{}^{5(s,t)}_{Jm},
\tilde{C}'{}^{(s,t)}_{Jm})^T$  
$(m=-J+1,-J+2,\cdots,J-1)$ 
satisfies (\ref{df=0}) with 
\begin{align}
A=
\left( 
\begin{array}{cccccc}
0 & -1 & 0 & 0 & 0 & 0 \\
\frac{3s}{c} & 2-4J(J+1) & 0 
& 0 & \frac{3\sqrt{2}}{c^2} & \frac{3\sqrt{2}m}{c^2}  \\
0 & 0 & \frac{s}{c}(1-2m) 
& 0 & -\sqrt{2}i \delta_- & -\sqrt{2}i \delta_- \\
0 & 0 & 0 
& \frac{s}{c}(1+2m) & -\sqrt{2}i\delta_+ & \sqrt{2}i\delta_+  \\
0 & 0 & \sqrt{2}i\delta_- 
& \sqrt{2}i\delta_+ & \frac{3s}{c} & \frac{2ms}{c} \\
0 & 0 & \sqrt{2}i\delta_- 
& -\sqrt{2}i\delta_+ & \frac{2ms}{c} & \frac{2s}{c} \\ 
\end{array}
\right).
\label{A for coker}
\end{align}
The eigenvalues of (\ref{A for coker}) do not satisfy (\ref{hantei}) and
hence, these modes do not contribute to the index.
On the other hand, the mode
$f=(C'{}^{(s,t)}_{JJ},d^{(s,t)}_{JJ}, \Upsilon^{-(s,t)}_{JJ-1},
\Upsilon'{}^{5(s,t)}_{JJ},\tilde{C}'{}^{(s,t)}_{JJ})^T$ 
satisfies (\ref{df=0}), where $A$ is given by $5\times 5$ matrix 
obtained by eliminating the row and column for $\Upsilon_+$ and 
putting $m=J$ in (\ref{A for coker}).
Then, we find that there is an eigenvalue which satisfies (\ref{hantei})
while the other four do not.
Therefore, together with the complex conjugate,
$f=(C'{}^{(s,t)}_{J-J},d^{(s,t)}_{J-J}, \Upsilon^{+(s,t)}_{J-J+1}
, \Upsilon'{}^{5(s,t)}_{J-J}, \tilde{C}'{}^{(s,t)}_{J-J})^T$, 
they contribute to the index. The eigenvalues of $R$ for 
these modes are given by $R \sim 2(\pm 2J + i(m_{si}-m_{tj}))$.
Note that this contribution is absent for $J=0$ because of 
the constraint (\ref{0 mode is 0}).
Finally, it is easy to see that $\Upsilon^{+(s,t)}_{J-J}$ and 
$\Upsilon^{-(s,t)}_{JJ}$ obey the equation 
$\partial \Upsilon +\frac{(2J+3)s}{c}\Upsilon =0 $. 
Then (\ref{hantei}) is satisfied for these modes, so that 
they contribute to the index. 
The value of $R$ for 
these are $R \sim 2(\pm (2J+2) + i(m_{si}-m_{tj}))$.

In summary, we find that
the determinant from the vector multiplet is given, up to an overall constants, by
\begin{align}
&\prod_{s,t=-\Lambda/2}^{\Lambda/2}
\prod_{\substack{J=|j_s-j_t| \\J\neq 0}}^{j_s+j_t}
\prod_{i=1}^{N_s}\prod_{j=1}^{N_t}
\{(2J)^2+(m_{si}-m_{tj})^2 \}^{1/2}
\nonumber\\
\times & \prod_{s,t=-\Lambda/2}^{\Lambda/2}\prod_{J=|j_s-j_t|}^{j_s+j_t}
\prod_{i=1}^{N_s}\prod_{j=1}^{N_t}
\{(2J+2)^2+(m_{si}-m_{tj})^2 \}^{1/2}.
\end{align}
Combined with (\ref{z hyper})
and the Vandermonde determinant
which comes from the diagonalization 
of the moduli matrix $M$,
the total one-loop determinant is given, up to an overall constant, by
\begin{align}
Z_{\rm 1-loop}=
\prod_{s,t=-\Lambda/2}^{\Lambda/2}
\prod_{J=|j_s-j_t|}^{j_s+j_t}
\prod_{i=1}^{N_s}\prod_{j=1}^{N_t}\hspace{-5.5mm} {\phantom{\prod}}^{\prime}
\left[
\frac{\{(2J+2)^2+(m_{si}-m_{tj})^2\} \{(2J)^2+(m_{si}-m_{tj})^2\}}
{\{(2J+1)^2+(m_{si}-m_{tj})^2\}^2}
\right]^{\frac{1}{2}}.
\label{1loopdet}
\end{align}
By $\prod'$ we mean that the second factor in the numerator with $s=t$, $J=0$ and $i=j$
is not included in the product.

\subsection{Partition function and Wilson loop}
The determinant (\ref{1loopdet}) depends on the background 
around which the theory has been expanded.
So it is labeled by a representation ${\cal R}$ of $SU(2)$.
If we ignore the instanton part,
the total partition function is written as
(\ref{sum of rep}), where $Z_{\cal R}$ is given by,
\begin{align}
&Z_{{\cal R}}= {\cal C}_{\cal R}\int \prod_{s=-\Lambda/2}^{\Lambda/2}
\prod_{i=1}^{N_s}dm_{si} Z_{\rm 1-loop}({\cal R},\{m_{si}\})
e^{-\frac{2}{g_{PW}^2}\sum_{s}\sum_{i}(2j_s+1)m_{si}^2}.
\label{matrix model}
\end{align}
The coefficient ${\cal C}_{\cal R}$ 
is a constant which determines the relative 
normalization in the sum (\ref{sum of rep}) and is given by
\begin{align}
\mathcal{C}_{\mathcal{R}}
=\prod_s \left(\frac{1}{2}\right)^{N_s^2} \cdot
\frac{N_{PW}!}{\prod_s \{N_s!(2j_s+1)!\}} \cdot
\prod_s\frac{(2\pi)^{N_s(N_s+1)/2}}{\prod_{k=1}^{N_s}k!}.
\end{align}
The first factor is the overall constant of $Z_{\text{1-loop}}$ we neglected.
The second factor is the number of ways to permute the eigenvalues of $L_4$
in the representation $\mathcal{R}$ and is part of the gauge volume.
The last factor is the product of the volume of $U(N_s)$, 
which arises from the diagonalization of $M_s$.
We ignore an overall constant which does not depend on the representation $\mathcal{R}$.
The Gaussian factor in (\ref{matrix model}) 
is obtained by substituting the saddle point 
configuration (\ref{QV=0}) to the original action of PWMM.


(\ref{matrix model}) for each representation ${\cal R}$ 
has a definite meaning. It describes the PWMM
expanded around the fuzzy sphere background with representation ${\cal R}$. 
Recall that the theories with $SU(2|4)$ symmetry are also realized as  
the theories around particular fuzzy sphere backgrounds in PWMM.
Then the partition functions of these theories can 
be obtained from (\ref{matrix model}) through the relations 
in Fig. 1 as we will see in the next section.

The Wilson loop (\ref{circle WL PWMM}) 
in PWMM is invariant under the supersymmetry 
(\ref{CKS in PWMM}). Hence, the calculation of its vev 
is also reduced to the matrix integral through the localization. 
At the saddle point, the operator (\ref{circle WL PWMM}) is reduced to 
\begin{align}
\frac{1}{N_{PW}}\sum_{s=-\Lambda/2}^{\Lambda/2}\sum_{i=1}^{N_s}
(2j_s+1)e^{2\pi m_{si}},
\label{WL in PWMM at SP}
\end{align}
where we used the fact that $M$ and $L_4$ commute and the eigenvalues of 
$L_4$ are integers or half-integers.
The vev of the Wilson loop is then reduced to 
the average of (\ref{WL in PWMM at SP})
with respect to the matrix integral (\ref{sum of rep}).

More generally, any operator in PWMM constructed only of $\phi$ is 
invariant under the supersymmetry with parameter (\ref{CKS in PWMM})
and hence its vev is reduced to a matrix integral.

In order to check our result, we perform the one-loop calculation of one-point
function of the operator 
${\rm Tr}\phi^2(0)={\rm Tr}(X_4+iX^{(E)}_0)^2(0)$ 
in the case of the trivial background,
based on two different method.
One is from the original action of PWMM and the 
other is from the eigenvalue integral (\ref{matrix model}).
We obtain the same results as shown in Appendix
\ref{Perturbative check}.


\section{Exact results for theories with $SU(2|4)$ symmetry}
In this section, we utilize the relations in Fig.~1 to 
obtain exact results for
SYM on $R\times S^2$ and SYM on $R \times S^3/Z_k$.
The partition functions of these theories can be 
obtained from (\ref{matrix model})
through the relations (a) and (c) in Fig.~1.
We also consider (c') and test the large-$N$ reduction for 
${\cal N}=4$ SYM on $R\times S^3$.

\subsection{SYM on $R\times S^2$}

Recall that the theory on $R\times S^2$ has many nontrivial vacua
in which gauge fields take the Dirac monopole configuration, 
(\ref{vacuum of SYM on RxS2}).
Each background is labeled by a set of monopole charges $\{q_s\}$
as well as their multiplicities $\{N_s\}$.
As shown in Section \ref{2+1 from PWMM}, 
the theory on $R\times S^2$ around each background is
realized from PWMM under the limit (\ref{S2 from PWMM}).
The partition function of the theory on $R\times S^2$ around 
the background labeled by
$\{(q_s,N_s) | s=-\Lambda/2, \cdots,\Lambda/2 \}$ is then 
obtained from (\ref{matrix model}) by taking the limit (\ref{S2 from PWMM});
\begin{align}
Z^{\{(q_s,N_s)\}}_{R\times S^2}=
&\int \prod_{s=-\Lambda/2}^{\Lambda/2} \prod_{i=1}^{N_s}dm_{si}
\prod_{s=-\Lambda/2}^{\Lambda/2}\Delta(m_s)^2
\prod_{s=-\Lambda/2}^{\Lambda/2} \prod_{i,j=1}^{N_s}
\left[
\frac{1+\left( \frac{m_{si}-m_{sj}}{2}  \right)^2}
{\{1+(m_{si}-m_{sj})^2\}^2}
\right]^{\frac{1}{2}}
\nonumber\\
&\prod_{s,t=-\Lambda/2}^{\Lambda/2}
\prod_{\substack{J=|q_s-q_t| \\ J\neq 0}}^{\infty}
\prod_{i=1}^{N_s}
\prod_{j=1}^{N_t}
\left[
\frac{\left\{1+\left(\frac{m_{si}-m_{tj}}{2J+2}\right)^2\right\} 
\left\{1+\left(\frac{m_{si}-m_{tj}}{2J}\right)^2\right\}}
{\left\{1+\left(\frac{m_{si}-m_{tj}}{2J+1}\right)^2\right\}^2}
\right]^{\frac{1}{2}}
e^{-\frac{8\pi }{g_{S^2}^2}\sum_{s,i} m_{si}^2},
\label{s2 nontrivial}
\end{align}
where we have dropped an overall constant and 
$\Delta(m_s)=\prod_{i<j}(m_{si}-m_{sj})$ is the Vandermonde 
determinant.
One can see that the infinite product of $J$ is convergent.
The full partition function is given as a sum over all 
(\ref{s2 nontrivial}) with $\{(q_s,N_s) \}$ satisfying 
$\sum_{s}N_s=N_{S^2}$.
Note that the commutative limit has been taken smoothly. 
This implies that the noncommutativity vanishes and does not
affect to the partition function, unlike the UV/IR mixing
on the Moyal plane.

For the trivial background given by $\Lambda=0$ and $q_0=0$ in 
(\ref{vacuum of SYM on RxS2}), 
(\ref{s2 nontrivial}) is simplified to
\begin{align}
Z^{\rm t.b.}_{R\times S^2}
=
&\int \prod_{i}dm_{i}
\prod_{i>j} \tanh^2 \left( 
\frac{\pi(m_i-m_j)}{2}
\right)
e^{-\frac{8\pi }{g_{S^2}^2}\sum_{i} m_{i}^2},
\end{align}
where $i,j$ run from $1$ to $N_{S^2}$.

The operator (\ref{WL in PWMM at SP}) is now reduced to 
\begin{align}
\frac{1}{N_{S^2}} \sum_{s=-\Lambda/2}^{\Lambda/2}
\sum_{i=1}^{N_s}e^{2\pi m_{si}}.
\end{align}
Through the relation (a) in Fig.~1, 
the vev of the above operator with respect to 
the matrix integral (\ref{s2 nontrivial}) is equal to 
the vev of (\ref{circle WL RxS2}) in SYM on $R\times S^2$ 
around the monopole background with $\{(q_s,N_s) \}$.

\subsection{${\cal N}=4$ SYM on $R\times S^3/Z_k$}

\subsubsection*{Taylor's T-duality}

SYM on $R\times S^3/Z_k$ is realized from PWMM through the 
relation (c) or (c') reviewed in Section \ref{3+1 from PWMM}.
We first apply the relation (c) which is based on Taylor's T-duality.
As explained in Section 2.2.3, $U(N)$
SYM on $R\times S^3/Z_k$ around trivial vacuum is obtained by
expanding PWMM around 
the background (\ref{vacuum of PWMM}) with $2j_s+1=n+ks$ and $N_s=N$
and imposing the orbifolding condition on the fluctuations. 
Applying this to (\ref{matrix model}) yields,
\begin{align}
Z^{\rm t.b.}_{R\times S^3/Z_k}
=& \int  \prod_{i=1}^N dm_i
\Delta(m)^2 \prod_{i,j=1}^{N}
\left[ 
\frac{1+\left(\frac{m_i-m_j}{2}\right)^2 }{\{1+(m_i-m_j)^2\}^2}
\right]^{\frac{1}{2}} 
\nonumber\\
&\prod_{u=-\infty}^{\infty}\prod_{\substack{J=|ku/2|\\J\neq 0}}^{\infty}\prod_{i,j=1}^{N}
\left[
\frac{\left\{ 1+\left(\frac{m_i-m_j}{2J+2} \right)^2   \right\}
\left\{ 1+\left(\frac{m_i-m_j}{2J}\right)^2  \right\}}
{\left\{1+\left(\frac{m_i-m_j}{2J+1}\right)^2\right\}^2}
\right]^{\frac{1}{2}}e^{-\frac{4\pi^2}{g^2}\sum_{i=1}^Nm^2_i},
\label{zktv}
\end{align} 
where $\Delta(m)=\prod_{i<j}(m_i-m_j)$ is the 
Vandermonde determinant and 
an over all constant is dropped.
The product of $u$ comes from the product of $s$ and $t$ in 
(\ref{matrix model}). Under the orbifolding condition, 
the blocks are labeled only by the difference, $u=s-t$, so that 
the only one product of $u$ is remaining in (\ref{zktv}).
The subscript $s$ of $m_{si}$ is also dropped for the same reason.
The exponent is obtained by using (\ref{renormalization for gpw and g}).

By changing the order of the products of $u$ and $J$ as 
\begin{align}
\prod_{u=-\infty}^{\infty}\prod_{\substack{J=|ku/2|\\J\neq 0}}^{\infty}=
\prod_{\substack{J\in Z/2 \\ J\geq 1/2}}
\prod_{\substack{u=-J \\ u\in kZ/2}}^{J},
\end{align}
one can first take the product of $u$ in (\ref{zktv}) for each $k$. 
For example, when $k=2$, the partition function becomes
\begin{align} 
Z^{\rm t.b.}_{R\times S^3/Z_2}
=&\int \prod_{i=1}^N dm_i
\Delta(m)^2 \prod_{i,j=1}^{N}
\left[ 
\frac{1+\left(\frac{m_i-m_j}{2}\right)^2 }{\{1+(m_i-m_j)^2\}^2}
\right]^{\frac{1}{2}} 
\nonumber\\
&
\prod_{J=1}^{\infty}\prod_{i,j=1}^{N}
\left[
\frac{\left\{ 1+\left(\frac{m_i-m_j}{2J+2} \right)^2   \right\}
\left\{ 1+\left(\frac{m_i-m_j}{2J}\right)^2  \right\}}
{\left\{1+\left(\frac{m_i-m_j}{2J+1}\right)^2\right\}^2}
\right]^{\frac{2J+1}{2}}
e^{-\frac{4\pi^2}{g^2}\sum_{i=1}^Nm^2_i},
\label{z2tv}
\end{align}
where, $J$ runs only over positive integers. 
One can see that the infinite product is convergent.

The theory on $R\times S^3/Z_k$ has nontrivial vacua 
labeled by a holonomy (\ref{holonomy}).
The partition functions of such theories are also obtained from 
(\ref{matrix model}) by taking appropriate representations shown in 
Section 2.2.3.

The vev of the circular Wilson loop operator 
(\ref{circle WL}) in $R\times S^3/Z_k$ 
with the contour (\ref{circular wl})
is reduced to the 
vev of the following operator with respect to the 
matrix integral (\ref{zktv}),
\begin{align}
\frac{1}{N} \sum_{i=1}^N e^{2\pi m_i}.
\end{align}
This operator is obtained from
(\ref{WL in PWMM at SP}) 
by dropping the $s$ dependence as above and using the formal expression
$N_{PW}=N n\Lambda $.

We then consider the case with $k=1$. In this case, 
${\cal N}=4$ SYM on $R\times S^3$ has a unique vacuum and 
its partition function is obtained in the same way as 
(\ref{z2tv}).
The partition function takes the same form as (\ref{z2tv})
except that $J$ runs over integers and half-integers starting from $1/2$.
Then it is easy to see that the measure factors
except the Vandermonde determinant completely cancel out.
Thus, we obtain the Gaussian matrix model. This is 
consistent with the results for ${\cal N}=4$ SYM obtained in 
\cite{Pestun:2007rz,Erickson:2000af,Drukker:2000rr}.

\subsubsection*{Large-$N$ reduction}
Alternatively, one can use the relation (c') in Fig.~1, which is 
based on the large-$N$ reduction, to obtain
the partition function of the planar SYM on $R\times S^3/Z_k$.
In particular, the theory around the trivial background is obtained 
from PWMM by taking the continuum limit (\ref{SYM on RxS3 from PWMM 2}) 
in PWMM around the background (\ref{SYM on RxS3 from PWMM 1}).
Applying this to (\ref{matrix model}), 
one can easily obtain the partition function.

In the following, we focus on the case of $k=1$ to check the claim of the 
large-$N$ reduction.  
In this case, before one takes the continuum limit,
the partition function is given by (\ref{matrix model})
with ${\cal R}$ given by (\ref{SYM on RxS3 from PWMM 1}) with $k=1$;
\begin{align}
Z^{\rm planar}_{R\times S^3}=
&\int \prod_{s=-\Lambda/2}^{\Lambda/2} \prod_{i,j=1}^{N}dm_{si}
\prod_{s,t=-\Lambda/2}^{\Lambda/2}
\prod_{J=|j_s-j_t|}^{j_s+j_t}
\prod_{i,j=1}^{N}\hspace{-5.5mm} {\phantom{\prod}}^{\prime}
\nonumber\\ 
&\left[
\frac{\{(2J+2)^2+(m_{si}-m_{tj})^2\} \{(2J)^2+(m_{si}-m_{tj})^2\}}
{\{(2J+1)^2+(m_{si}-m_{tj})^2\}^2}
\right]^{\frac{1}{2}}e^{-\frac{2}{g_{PW}^2}\sum_{s,i}(n+s) m_{si}^2},
\label{n=4symfromPWMM}
\end{align}
where $2j_s+1=n+s$. We show that in the continuum limit 
(\ref{SYM on RxS3 from PWMM 2}), 
this matrix integral is indeed equivalent to the Gaussian matrix 
model (\ref{GMM}) of ${\cal N}=4$ SYM. 
Since $n \gg s $ for $s=-\Lambda/2,\cdots,\Lambda/2$ in the continuum limit, 
the exponent in (\ref{n=4symfromPWMM}) goes to 
$-\frac{4\pi^2}{g^2}\sum_{s,i}m_{si}^2$, where the identification 
for the coupling constants in (\ref{SYM on RxS3 from PWMM 1}) is used.
The coefficient in the exponent agrees with that in (\ref{GMM}).
We will see that the interactions between the modes with 
different $s$ in (\ref{n=4symfromPWMM}) are suppressed in the 
continuum limit 
and the model becomes a set of independent copies of 
the Gaussian matrix model.

We assume the 't Hooft limit in the following. 
Then the saddle point approximation is exact. 
We introduce the eigenvalue density for each $s$ as 
\begin{align}
\rho^{[s]}(x)=\frac{1}{N}\sum_{i=1}^{N}\delta (x-m_{si}).
\end{align}
The saddle point equation for $\rho^{[s]}$ is given by
\begin{align}
0= \frac{2}{\lambda_s}x 
-\sum_{t=-\Lambda/2}^{\Lambda/2} 
\sum_{J=|j_s-j_t|}^{j_s+j_t}
& \int dy \rho^{[t]}(y)(x-y)
\left\{ 
\frac{1}{(2J+2)^2+(x-y)^2}
\right.
\nonumber\\ 
& \left. +
\frac{1}{(2J)^2+(x-y)^2}
-\frac{2}{(2J+1)^2+(x-y)^2}
\right\},
\label{spe}
\end{align}
where $\lambda_s=g_{PW}^2N/(n+s)$.
One can rewrite (\ref{spe}) to
\begin{align}
0=\; &\frac{2}{\lambda_s}x-\int dy \frac{\rho^{[s]}(y)}{x-y}
+2\sum_{t=-\Lambda/2}^{\Lambda/2}
\sum_{J=|j_s-j_t|}^{j_s+j_t}
f_J^{[t]}(x)
\nonumber\\
&-\sum_{t=-\Lambda/2-1}^{\Lambda/2-1}
\sum_{J=|j_s-j_t|}^{j_s+j_t}
f_J^{[t+1]}(x)
-\sum_{t=-\Lambda/2+1}^{\Lambda/2+1}
\sum_{J=|j_s-j_t|}^{j_s+j_t}
f_J^{[t-1]}(x)
\label{spe2}
\end{align}
where $f_J^{[t]}(x)$ is defined by
\begin{align}
f_J^{[t]}(x)=\int dy \rho^{[t]}(y)\frac{x-y}{(2J+1)^2+(x-y)^2},
\end{align}
for $J=0,1,2,\cdots$ and $t=-\Lambda/2, \cdots, \Lambda/2$.

In the continuum limit (\ref{SYM on RxS3 from PWMM 2}), 
the saddle point equation (\ref{spe2}) is solved by 
\begin{align}
\rho^{[s]}(x)=\hat{\rho}(x):=\frac{2}{\lambda}\sqrt{\lambda-x^2},
\label{semi circle}
\end{align}
for any $s \in Z$. Here $\lambda=g^2N/(2\pi^2)$, which is the limit of $\lambda_s$
under the identification \eqref{SYM on RxS3 from PWMM 2}.
The distribution of $\hat{\rho}$ 
is just the semicircle law and we find that 
the model consists of infinitely many copies of 
the Gaussian matrix model in the continuum limit.

One can see the equivalence for some physical observables.
For example, the free energy of the reduced model, \eqref{n=4symfromPWMM},
divided by the multiplicity $\Lambda$ is 
equal to that of ${\cal N}=4$ SYM,
\begin{align}
\frac{F_{r}}{\Lambda}=F_{{\cal N}=4 \; {\rm SYM}},
\end{align}
in the continuum limit. 
The left-hand side can be computed by using (\ref{semi circle})
and the right-hand side is the free energy of the Gaussian matrix model
in (\ref{GMM}).
Also the VEV of the circular Wilson loop (\ref{WL in PWMM at SP}) 
is calculated in (\ref{n=4symfromPWMM}) as
\begin{align}
\frac{1}{\Lambda N} \sum_{s}\sum _{i} 
\langle e^{2\pi m_{si}}
\rangle
=
\frac{1}{\Lambda}
\sum_{s}
\int dx \rho^{[s]}(x) e^{2\pi x}
=\int dx \hat{\rho}(x) e^{2\pi x},
\label{wllp}
\end{align}
where we have used the relation $N_{PW} \sim n \Lambda N $ which 
holds in the continuum limit. 
The right-hand side of (\ref{wllp})
is nothing but the known result in 
${\cal N}=4$ SYM, (\ref{GMM}). 

In the above argument, we have ignored a cutoff effect.
When $s$ is sufficiently close to the cutoff $\pm \Lambda/2$,
$\rho^{[s]}$ should deviate from 
the semicircle law (\ref{semi circle}) since the last three terms in 
(\ref{spe2}) do not vanish with (\ref{semi circle}). 
However, this deviation rapidly disappears when $s$ goes to a distance from 
the cutoffs. More precisely, when $|s|<\Lambda/2 -{\cal O}(\log \Lambda)$, 
the cutoff effect in (\ref{spe2}) can be neglected. 
In fact, the cutoff effect is caused by the terms with 
$t={\cal O}(\Lambda)$ in the last three terms in (\ref{spe2}) and when 
$|s|<\Lambda/2 -{\cal O}(\log \Lambda)$ such effect is suppressed 
since the lower edge of $J$ is at least ${\cal O}(\log \Lambda)$. 
Hence the deviation from (\ref{semi circle}) appears
only when the distance from $s$ to the cutoff is ${\cal O}(1)$.
This means that the number of the deviating modes is 
${\cal O}(1)$ and it is negligible compared with the total number 
of the modes, $\Lambda+1$, and then the other modes satisfying 
(\ref{semi circle}) are dominant in the continuum limit.
Since, as we have seen above, an expectation value in the reduced model
is written as an average over all the modes, 
contribution from the deviating modes are suppressed\footnote{
Similar situations are found also in the large-$N$ reduced model 
for Chern-Simons theories on $S^3$
\cite{Ishiki:2009vr,Ishiki:2010pe,Asano:2012gt}.}.

In \cite{Ishiki:2011ct}, 
the large-$N$ equivalence for the circular Wilson loop was studied 
in the perturbation theory.
Within the ladder approximation, it was shown that 
the vev of (\ref{circle WL PWMM}) agrees with (\ref{GMM})
to all orders in the perturbative expansion.
The above result provides a nonperturbative proof of 
the large-$N$ equivalence for the free energy and the 
circular Wilson loop operator.

\section{Summary}
In this paper, we used the localization technique 
to obtain the matrix integral \eqref{matrix model}, which is
equivalent to the partition function of PWMM around the fuzzy sphere vacuum 
with the representation $\mathcal{R}$.
We first constructed off-shell supersymmetries in PWMM 
and added a $Q$-exact term to the action.
Then, the path integral is reduced to the one-loop integral around saddle points 
of the $Q$-exact term. 
Except for possible instanton effects,
the saddle points are given by fuzzy spheres labeled by an $SU(2)$ representation.
In the end, up to the instantons, the partition function 
is given by a sum of terms, each of which is labeled by an $SU(2)$ representation and 
given by a matrix integral.
We also obtained the vev of $Q$-closed operators as the matrix integral.
As a consistency check of our results,
we performed one-loop computation in PWMM
and found the exact agreement with the result obtained by using the localization.
Although the instanton effects are not included in our computation,
in the 't Hooft limit, where the instanton effects are negligible, our results are exact.

Using the relations (a) and (c) explained in Section 2.2.1 and 2.2.3,
we obtained matrix integrals equivalent to the partition function of
theories with $SU(2|4)$ symmetry, 2+1 SYM on $R\times S^2$ and $\cN=4$ SYM on $R\times S^3/Z_k$.
The $SU(2|4)$ symmetric theories have many nontrivial vacua.
The theory around each vacuum of these theories are realized by PWMM around a particular 
fuzzy sphere vacuum through (a) and (c).
We applied these relations to \eqref{matrix model}
and obtained the matrix integral for $SU(2|4)$ symmetric theories.
In the case of $\cN=4$ SYM on $R\times S^3$, 
we saw that our result correctly reproduces 
the Gaussian matrix model of $\cN=4$ SYM \cite{Pestun:2007rz,Erickson:2000af,Drukker:2000rr}.

We also considered the relation (c') in Fig. 1. 
This is regarded as the large-$N$ reduction for theories on 
$R\times S^3/Z_k$.
From the result of the localization, we obtained the partition 
function and the vev of the circular Wilson loop 
in the reduced model of SYM on $R\times S^3$.
We found that the free energy and the circular Wilson loop agree 
between the reduced model and SYM on $R\times S^3$.
Our result provides a non-perturbative proof of the 
large-$N$ equivalence for these observables.

It may be possible to compute the matrix integral \eqref{matrix model} exactly.
If not, at least, one can compute it numerically.
It is interesting to compare these results with gravity duals.
The gravity dual of \eqref{matrix model} for each $\mathcal{R}$ is constructed 
in \cite{Lin:2005nh}. 
It would be possible to study the gauge/gravity duality
for the family of various different theories labelled by 
$\mathcal{R}$ in a unified manner.

The remaining task to obtain the full partition function of PWMM is 
to compute the instanton part, which we have not addressed in this paper.
As noted in the last part of Section \ref{section qv}, 
the saddle point equations at the future and the past infinities
reduce to anti-self-dual and self-dual equations, respectively.
They are indeed the mass deformed Nahm equations \cite{Bachas:2000dx}.
By examining the moduli space of these equations,
it would be possible to obtain the instanton corrections to our results,
which may shed light on the nature of M-theory.
We hope to report on these issues in the near future.

\section*{Acknowledgements}
The work of Y.A. is supported by the Grant-in-Aid for the Global COE Program 
``The Next Generation of Physics, Spun from Universality and Emergence'' 
from the Ministry of Education, Culture, Sports, Science and Technology (MEXT) of Japan. 
The work of G.I. T.O. and S.S. is supported 
in part by the JSPS Research Fellowship for Young Scientists.

\appendix

\section{Gamma matrices}
Our gamma matrices are the same as those in \cite{Pestun:2007rz}.

The local Lorentz metric is ``mostly plus'', 
$g_{MN}={\rm diag}(-1,1,1,\cdots,1)$ $(M,N=0,1,\cdots,9)$.
The ten-dimensional $32 \times 32$ gamma matrices $\gamma^M$ ($M=0,1,\cdots,9$) obey
\be
\gamma^{\{M} \gamma^{N\}}= g^{MN}.
\ee
The associated representation of $Spin(1,9)$ can be decomposed into two irreducible representations by the chirality,
\be
\gamma^{11}\equiv \gamma^1 \cdots \gamma^9 \gamma^0.
\ee

We decompose the ten-dimensional  Dirac spinor as 
\be
\begin{pmatrix}
S^+ \\
S^-
\end{pmatrix}.
\ee
Then, the gamma matrices $\gamma^M$ are expressed in the block form, 
\be
\gamma^M=\begin{pmatrix}
0& \tilde{\Gamma}^M\\
\Gamma^M & 0
\end{pmatrix}.
\ee
We take $\Gamma^M,\ \tilde{\Gamma}^M$ to be symmetric;
\be
(\Gamma^M)^{T}= \Gamma^M,\ (\tilde{\Gamma}^M)^T =\tilde{\Gamma}^M.
\ee
We define $\gamma^{MN},\ \Gamma^{MN}$, and  $\tilde{\Gamma}^{MN}$ as 
\bea
\gamma^{MN}\equiv \gamma^{[M}\gamma^{N]}=
\begin{pmatrix}
\tilde{\Gamma}^{[M}\Gamma^{N]}&0\\
0& \Gamma^{[M}\tilde{\Gamma}^{N]}
\end{pmatrix}\equiv \begin{pmatrix}
\Gamma^{MN}&0 \\
0 & \tilde{\Gamma}^{MN}
\end{pmatrix}.
\eea
Then, we have
\bea
\tilde{\Gamma}^{\{ M}\Gamma^{N\}}&=&\Gamma^{\{M}\tilde{\Gamma}^{ N\}}=g^{MN}, \\
\Gamma^M \Gamma^{PQ}&=& 4 g^{M[P}\Gamma^{Q]} + \tilde{\Gamma}^{PQ}\Gamma^M.
\eea
We  write some useful identities:
\bea
(\Gamma_M)_{\alpha_1 \{ \alpha_2} (\Gamma^M)_{\alpha_3 \alpha_4 \}}&=&0,
\label{triality}\\
(\Gamma^M)_{\alpha \delta} (\Gamma_M)_{\gamma \beta}&=& -\frac{1}{2} (\Gamma^M)_{\alpha \beta}(\Gamma_M)_{\gamma \delta} +\frac{1}{24} (\Gamma^{MNP})_{\alpha \beta} (\Gamma_{MNP})_{\gamma \delta},\\
(\tilde{\Gamma}^{MN})_\alpha^{\ \beta} (\tilde{\Gamma}_{MN})_\gamma^{\ \delta}&=&4 (\Gamma^M)_{\alpha \gamma}(\tilde{\Gamma}_M)^{\beta \delta} -2 \, \delta_{\alpha}^\beta \delta_\gamma^\delta -8\, 
\delta_\alpha^\delta \delta_\gamma^{\beta},
\eea
where  $\alpha, \beta,\cdots$ are spinor indices. The first equality  is so called ``triality'', and the last two are Fierz identities. 

Decomposing the indices $M=0,\cdots,9$ into $a=1,2,3,4$ and $m=0,5,\cdots,9$,  we obtain the following identities
\bea
\Gamma_{a m}\tilde{\Gamma}^{a}&=& -4 \tilde{\Gamma}_m,\\
\Gamma^a \Gamma_{b c}\tilde{\Gamma}_a&=&0,\\
\Gamma^{a}\Gamma_{b m}\tilde{\Gamma}_{a}&=&2\tilde{\Gamma}_{bm},\\
\Gamma^{a}\Gamma_{mn}\tilde{\Gamma}_a &=& 4 \tilde{\Gamma}_{mn}.
\eea

In the rest of this appendix, we write down the gamma matrices $\Gamma^M$ and $\tilde{\Gamma}^M$ explicitly.
\bea
\Gamma^0 &=& \begin{pmatrix}
1_{8\times8} &0\\
0 & 1_{8\times8}\\
\end{pmatrix},\ \
 \Gamma^9 = \begin{pmatrix}
1_{8\times8} &0\\
  0 & -1_{8\times8} 
\end{pmatrix},\\
\Gamma^i &=& \begin{pmatrix}
0 &E^T_i\\
E_i & 0 
\end{pmatrix} \ (i=1,\cdots,8).\nonumber
\eea
The $8\times 8$ matrices $E_i\ ( i=1,\cdots,8)$ are given by
\bea
E_a = \begin{pmatrix}
J_a & 0 \\
0 & \bar{J}_a
\end{pmatrix} \ \ (a=1,2,3,4),\ \ E_{m'} = \begin{pmatrix}
0 & -J_{m'}^T \\
J_{m'} & 0
\end{pmatrix}\ \ (m=5,6,7,8).
\eea 
Finally, the $4\times4$ matrices $J_a, \bar{J}_a$  are given as follows;
\bea
&J_1 = 1_{4\times 4},\  \bar{J}_1=1_{4\times4},\nonumber\\
 &J_2 = \begin{pmatrix}
0&-1&0&0\\
1&0&0&0\\
0&0&0&-1\\
0&0&1&0 
\end{pmatrix},\ 
J_3 = \begin{pmatrix}
0&0&-1&0\\
0&0&0&1\\
1&0&0&0\\
0&-1&0&0
 \end{pmatrix},\ 
 J_4 = \begin{pmatrix}
0&0&0&-1\\
0&0&-1&0\\
0&1&0&0\\
1&0&0&0
 \end{pmatrix},\\
&\bar{J}_2 = \begin{pmatrix}
0&-1&0&0\\
1&0&0&0\\
0&0&0&1\\
0&0&-1&0 
\end{pmatrix},\
\bar{J}_3 = \begin{pmatrix}
0&0&-1&0\\
0&0&0&-1\\
1&0&0&0\\
0&1&0&0
 \end{pmatrix},\
 \bar{J}_4 = \begin{pmatrix}
0&0&0&-1\\
0&0&1&0\\
0&-1&0&0\\
1&0&0&0
 \end{pmatrix},\nonumber
\eea
and the matrices $J_m$ are given by
\bea
J_ 5= \begin{pmatrix}
1&0&0&0\\
0&-1&0&0\\
0&0&-1&0\\
0&0&0&-1 
\end{pmatrix},
J_6 = \begin{pmatrix}
0&1&0&0\\
1&0&0&0\\
0&0&0&1\\
0&0&-1&0 
\end{pmatrix},\\
J_7 = \begin{pmatrix}
0&0&1&0\\
0&0&0&-1\\
1&0&0&0\\
0&-1&0&0
 \end{pmatrix},\ 
 J_8 = \begin{pmatrix}
0&0&0&1\\
0&0&1&0\\
0&-1&0&0\\
1&0&0&0
 \end{pmatrix}. \nonumber
\eea
The matrices  $J_{a'}$ and $\bar{J}_{a'}$ satisfy
\bea
J_{a'} J_{b'}= -\delta_{a'b'}{\bf 1}_{4}+\varepsilon_{a' b' c'} J_{c'},\ \ 
\bar{J}_{a'} \bar{J}_{b'}=-\delta_{a'b'}{\bf 1}_{4} -\varepsilon_{a' b' c'} \bar{J}_{c'} \ \ (a',\, b' ,\ c' = 2,3,4).
\eea

Note that, in this representation,  we have
\bea
\Gamma^{1234}&=&\Gamma^{1}\Gamma^2 \Gamma^3 \Gamma^4 = \begin{pmatrix}
1_{4\times4}& 0&0&0\\
0&-1_{4\times4}& 0 &0\\
0&0&-1_{4\times4}&0\\
0&0&0&1_{4\times4}
\end{pmatrix},\\ \ \Gamma^{5678}&=&\Gamma^{5}\Gamma^6 \Gamma^7 \Gamma^8 = \begin{pmatrix}
1_{4\times4}& 0&0&0\\
0&-1_{4\times4}& 0 &0\\
0&0&1_{4\times4}&0\\
0&0&0&-1_{4\times4}
\end{pmatrix}.\nonumber
\eea

\section{Our convention for $S^3$}\label{s3}

In this appendix, we summarize our convention for $S^3$ with a unit 
radius (See also \cite{Ishii:2008tm,Ishii:2008ib}).
$S^3$ is viewed as the $SU(2)$ group manifold. We parametrize an element
of $SU(2)$ in terms of the Euler angles as
\begin{equation}
g=e^{-i\varphi \mathcal{J}_4/2}e^{-i\theta \mathcal{J}_3/2}e^{-i\psi \mathcal{J}_4/2},
\label{Euler angles}
\end{equation}
where $0\leq \theta\leq \pi$, $0\leq \varphi < 2\pi$, $0\leq \psi < 4\pi$ 
and $\mathcal{J}_{a'}\ (a'=2,3,4)$ satisfies $[\mathcal{J}_{a'}, 
\mathcal{J}_{b'}]=i \varepsilon_{a' b' c'}\mathcal{J}_{c'}$. 
The periodicity for these angle variables is given by
\begin{align}
(\theta,\varphi,\psi)\sim (\theta,\varphi+2\pi,\psi+2\pi)\sim (\theta,\varphi,\psi+4\pi).
\label{periodicity on S^3}
\end{align}

The isometry of $S^3$ corresponds to the left and the right multiplications
of $SU(2)$ elements on $g$. 
We construct the right-invariant 1-forms under the multiplications,
\begin{equation}
dgg^{-1}=-i e^{a' }\mathcal{J}_{a'}.
\label{ri 1-form}
\end{equation}
The explicit form of $e^{a'}$ is given by
\begin{eqnarray}
&&e^2=\frac{1}{2}(-\sin \varphi d\theta + \sin\theta\cos\varphi d\psi),\nonumber\\
&&e^3=\frac{1}{2}(\cos \varphi d\theta + \sin\theta\sin\varphi
 d\psi),\nonumber\\
&&e^4=\frac{1}{2}(d\varphi + \cos\theta d\psi).
\label{ri explicit}
\end{eqnarray} 
It is easy to see that $e_{a'}$ satisfy the Maurer-Cartan equation,
\begin{equation}
de^{a'}-\varepsilon_{a' b' c' }e^{b'}\wedge e^{c'}=0.\label{Maurer-Cartan}
\end{equation}
We take $e^{a'}$ as the vielbein in this paper. In this frame, 
the spin connection is simply given by $\omega^{a' b'}=\varepsilon^{a'b'c'}e^{c'}$.
The metric is given by
\begin{equation}
ds^2=e^{a'}e^{a'}=\frac{1}{4}\left(
d\theta^2+\sin^2\theta d\varphi^2 +(d\psi+\cos\theta d\varphi)^2\right).
\label{metric of S^3}
\end{equation}

The Killing vectors ${\cal{L}}_{a'}$ dual to $e^{a'}$ are given by
\begin{equation}
{\cal{L}}_{a'}=-\frac{i}{2}e^\mu_{a'}\partial_\mu,
\end{equation}
where $\mu=\theta,\varphi,\psi$, and $e^\mu_{a'}$ are inverse of $e^{a'}_\mu$. The explicit form of the Killing
vectors are
\begin{eqnarray}
&&{\cal{L}}_2=-i\left(-\sin\varphi\partial_{\theta}-\cot\theta\cos\varphi\partial_{\varphi}+\frac{\cos\varphi}{\sin\theta}\partial_{\psi}\right),\nonumber\\
&&{\cal{L}}_3=-i\left(\cos\varphi\partial_{\theta}-\cot\theta\sin\varphi\partial_{\varphi}+\frac{\sin\varphi}{\sin\theta}\partial_{\psi}\right),\nonumber\\
&&{\cal{L}}_4=-i\partial_{\varphi}.\label{Killing vector}
\end{eqnarray}
Because of the Maurer-Cartan equation (\ref{Maurer-Cartan}), the Killing vectors satisfy the SU(2) algebra, $[{\cal{L}}_{a'},{\cal{L}}_{b'}]=i\varepsilon_{a' b' c'}{\cal{L}}_{c'}$.

\section{Monopole spherical harmonics}

Here, we write down the monopole spherical harmonics \cite{Ishii:2008tm}.
One can regard $S^3$ as a $U(1)$ bundle over $S^2=SU(2)/U(1)$.
$S^2$ is parametrized by $\theta$ and $\varphi$ and covered with two local patches:
the patch I defined by $0\leq\theta <\pi$ and the patch II defined by $0<\theta\leq\pi$.
In the following expressions, the upper sign is taken in the patch I while the lower sign in the patch II.
The element of $SU(2)$ in (\ref{Euler angles}) is decomposed as 
\begin{align}
&g=L\cdot h \;\;
\mbox{with}\;\;
L=e^{-i\varphi \mathcal{J}_4/2}e^{-i\theta \mathcal{J}_3/2}e^{\pm i\varphi \mathcal{J}_4/2}\;\;\mbox{and}\;\;
h=e^{-i(\psi\pm\varphi) \mathcal{J}_4/2}.
\end{align}
$L$ is a representative of $SU(2)/U(1)$,
while $h$ represents the fiber $U(1)$.
The fiber direction is parametrized by $y=\psi\pm\varphi$.
Note that $L$ has no $\varphi$-dependence for $\theta=0,\pi$.
The zweibein of $S^2$ is given by the $a'=2,3$ components of the left-invariant 1-form, 
$-iL^{-1}dL=2  \ e^{a'} {\mathcal{J}_{a'}}/{2}$ \cite{Salam:1981xd}.
It takes the form
\begin{align}
&e^{2}=\frac{1}{2}(\pm \sin\varphi d \theta+\sin \theta\cos\varphi d\varphi), \nonumber
&e^{3}=\frac{1}{2}(-\cos\varphi d \theta \pm \sin\theta\sin\varphi d\varphi).
\end{align}
This zweibein gives the standard metric of $S^2$ with the radius $1/2$:
\begin{align}
ds^2=\frac{1}{4}(d\theta^2+\sin^2\theta\varphi^2).
\label{metric of S^2}
\end{align}
Making a replacement $\partial_y \rightarrow -iq$ in (\ref{Killing vector})
leads to the angular momentum operator in the presence of a monopole with
magnetic charge $q$ at the origin \cite{Wu:1976ge}:
\begin{eqnarray}
&&L_2^{(q)}=i(\sin\varphi\partial_{\theta}+
\cot\theta\cos\varphi\partial_{\varphi})-
q\frac{1\mp \cos\theta}{\sin\theta}\cos\varphi, \nonumber\\
&&L_3^{(q)}=i(-\cos\varphi\partial_{\theta}+
\cot\theta\sin\varphi\partial_{\varphi})-
q\frac{1\mp \cos\theta}{\sin\theta}\sin\varphi, \nonumber\\
&&L_4^{(q)}=-i\partial_{\varphi}\mp q ,
\label{monopole angular momentum}
\end{eqnarray}
where $q$ is quantized as 
$q=0, \pm \frac{1}{2}, \pm 1, \pm \frac{3}{2},\cdots$,
because $y$ is a periodic variable with the period $4\pi$.
These operators act on the local sections on $S^2$ and satisfy the $SU(2)$
algebra $[L_{a'}^{(q)},L_{b'}^{(q)}]=i\varepsilon_{a' b' c'}L_{c'}^{(q)} $. 
Note that when $q=0$, these operators are reduced to \eqref{Killing vector} with $\partial_\psi=0$, which is  the ordinary angular momentum operators  on $S^2$.
The $SU(2)$ acting on $g$ from left survives as the isometry of $S^2$.
Note that in 2+1 SYM on $R\times S^2$ the isometry of $S^2$ is included
in the $SU(2|4)$ symmetry as a subgroup.

The monopole spherical harmonics are the basis of local sections on
$S^2$. They are given by
\begin{equation}
\tilde{Y}_{Jmq}(\Omega_2)=
(-1)^{J-q}\sqrt{2J+1}\langle J -q|e^{i\theta \mathcal{J}_3}|Jm\rangle
e^{i(\pm q+m)\varphi}.
\label{monopole harmonics}
\end{equation}
Here, $J=|q|,|q|+1,|q+2|,\cdots$, $m=-J,-J+1,\cdots , J-1,J$.
The existence of the lower bound of the angular momentum $J\geq|q|$ 
is due to the fact that the magnetic field produced by the 
monopole also has nonzero angular momentum.  
Note that the monopole harmonics with $q=0$ do not transform
on the overlap of two patches. They correspond to 
global sections (functions) on $S^2$ which are expressed by
the ordinary spherical harmonics on $S^2$.
The action of $L^{(q)}_{a'}$ on the monopole spherical harmonics 
is given by
\begin{eqnarray}
(L^{(q)})^2\tilde{Y}_{Jmq}&=&J(J+1)\tilde{Y}_{Jmq},\nonumber\\
L^{(q)}_{\pm}\tilde{Y}_{Jmq}&=&\sqrt{(J\mp m)(J \pm
 m+1)}\tilde{Y}_{Jm\pm 1q},\nonumber\\
L^{(q)}_4\tilde{Y}_{Jmq}&=&m\tilde{Y}_{Jmq},
\end{eqnarray}
where $L^{(q)}_{\pm}\equiv L^{(q)}_{2}\pm i L^{(q)}_{3}$.  The complex conjugates of the monopole spherical harmonics obeys the following relation,
\begin{equation}
\left(\tilde{Y}_{Jmq}\right)^*=(-1)^{m-q}\tilde{Y}_{J-m-q}.
\end{equation}
The monopole spherical harmonics are orthonormal to each other;
\begin{equation}
\int \frac{d\Omega_2}{4\pi}\left(\tilde{Y}_{Jmq}\right)^*
\tilde{Y}_{J'm'q}=\delta_{JJ'}\delta_{mm'}.
\end{equation}

\section{Fuzzy spherical harmonics}\label{A.harmonics}
In this appendix, we review the fuzzy spherical harmonics which
form a basis of rectangular matrices \cite{Ishii:2008ib,Ishiki:2006yr}.

Let us consider a $(2j_s+1)\times (2j_t+1)$ rectangular complex matrix,
where $j_s,j_t \in Z_{\geq 0}/2$.
Such a matrix $M^{(s,t)}$ can be generally expanded as
\begin{equation}
M^{(s,t)}=\sum_{m_s,m_t}M_{m_sm_t}|j_sm_s\rangle \langle j_tm_t |,
\end{equation}
by using a basis $\{ |jm \rangle \;|\ m=-j,-j+1,\cdots,j \}$ 
of the spin $j$ representation space 
of $SU(2)$ algebra. 
We define an operation which multiplies the representation 
matrices of the $SU(2)$ generators from left and right:
\begin{align}
L_{a'}\circ M^{(s,t)}&=\sum_{m_s,m_t}M_{m_sm_t}(L_{a'}^{[j_s]}|j_sm_s\rangle \langle j_tm_t |-|j_sm_s\rangle \langle j_tm_t |L_{a'}^{[j_t]}),
\label{LcM}
\end{align}
where $L_{a'}^{[j]}\ (a'= 2,3,4)$ stands for the spin $j$ representation matrix of the 
generator.

We can construct another basis of the rectangular matrices 
denoted by $\{ \hat Y_{Jm(j_sj_t)} \}$ such that they satisfy
\begin{align}
(L_{a'}\circ )^2\hat Y_{Jm(j_sj_t)}=&J(J+1)\hat Y_{Jm(j_s,j_t)},\nonumber \\
L_\pm \circ \hat Y_{Jm(j_sj_t)}=&\sqrt{(J\mp m)(J\pm m+1)}\hat Y_{Jm\pm 1(j_s,j_t)},\nonumber \\
L_4\circ \hat Y_{Jm(j_sj_t)}=&m\hat Y_{Jm(j_s,j_t)}.
\end{align}
$ \hat Y_{Jm(j_sj_t)}$ are
called scalar fuzzy spherical harmonics and defined by
\begin{equation}
\hat Y_{Jm(j_sj_t)}=
\sum_{m_s,m_t}(-)^{-j_s+m_t}C^{Jm}_{j_sm_sj_tm_t}|j_sm_s\rangle 
\langle j_tm_t|,
\end{equation}
where $C^{Jm}_{j_sm_sj_tm_t}$ are the 
Clebsch-Gordan coefficients.
Their hermitian conjugates are given by
\begin{equation}
(\hat Y_{Jm(j_sj_t)})^\dagger =(-)^{m-(j_s-j_t)}\hat Y_{J-m(j_tj_s)},
\end{equation}
and they satisfy the orthogonality relation
\begin{equation}
\tr \left\{ (\hat Y_{Jm(j_sj_t)})^\dagger 
\hat Y_{J^\prime m^\prime (j_s  j_t )}\right\} =\delta _{J,J^\prime}\delta _{m,m^\prime}.
\end{equation}

\section{Perturbative check of our result for  trivial background }
\label{Perturbative check}
We consider the following observable  around the trivial background, $\hat{X}_{a'}=0$, in the PWMM, 
\be
\langle {\rm Tr}\,( X_4 + i X^{(E)}_0 )^2 (\tau=0) \rangle. \label{eq:PWMMobs}
\ee
This observable is $Q$-closed and so can be computed by the localization method.
 In this appendix, in order to illustrate the validity of the matrix integral \eqref{matrix model}, we  will compute this observable perturbatively both from the original PWMM and the matrix integral \eqref{matrix model}. 
We will see that the two different computations agree completely 
up to the one-loop level.

\subsection*{One-loop calculation in PWMM}\label{sec:1loopPWMM}
In the trivial background, the action of PWMM $S=S_{free}+S_{int}$ is given by
\bea
S_{free}&=&\frac{1}{g^2_{PW}}\int d\tau{\rm Tr}\left[ -\frac{1}{2}( \partial_\tau X_{a'})^2 -2 X_{a'}^2 -\frac{1}{2} X_m^2 -\frac{i}{2} \Psi \Gamma^1 \partial_1 \Psi -\frac{3i}{4} \Psi \Gamma^{234}\Psi \right],\\
S_{int}&=&\frac{1}{g^2_{PW}}\int d\tau{\rm Tr}\biggl[ i\varepsilon_{a' b' c'} X_{a'}[X_{b'}, X_{c'}] +\frac{1}{4} [X_{a'}, X_{b'}]^2 +\frac{1}{2}[ X_{a'},X_{m}]^2 +\frac{1}{4}[X_m,X_n ]^2\nonumber\\ 
&&\ \ \ \ \ \ \ \ \ \ \ \ \ \ \ \ \ \ -\frac{1}{2}\Psi \Gamma^M [X_M, \Psi] \biggr]. \label{eq:int trivial}
\eea
Here, we have taken $X_1=0$ gauge, and $M=0,1,\cdots,9$,  $a'=2,3,4$ and $m=5,\cdots,9$. 

 We can read off the Feynman rule of PWMM in momentum space. The propagators are given by 
\bea
\langle X_{a', ij}(p) X_{b', kl}(q) \rangle&=& 2 \pi \delta(p+q)\delta_{a' b'}\delta_{il}\delta_{jk}\frac{g^2_{PW}}{p^2+4},\\
\langle X_{m, ij}(p) X_{n, kl}(q) \rangle&=& 2 \pi \delta(p+q)\delta_{mn }\delta_{il}\delta_{jk}\frac{g^2_{PW}}{p^2+1}, \label{X0prop}\\
\langle \Psi_{\alpha, ij}(p) \Psi_{\beta, kl}(q) \rangle&=& 2 \pi \delta(p+q)\delta_{il}\delta_{jk}\frac{ (p \Gamma^1+ \frac{3 i }{2} \Gamma^{234})_{\alpha\beta}}{p^2 + \frac{9}{4}}g^2_{PW}. 
\eea
Note that \eqref{X0prop} with $m=n=0$ is not the propagator of $X_0$, but that of the wick rotated field $X_0^{(E)}$.

We compute \eqref{eq:PWMMobs} up to the one-loop order. 
Note that the term $\langle  {\rm Tr} \, X_4 X_0^{(E)} \rangle$ vanishes 
up to the one-loop level, we compute
\bea
\langle {\rm Tr} \,X_4 X_4 \rangle - \langle  {\rm Tr} \, X_0^{(E)} 
X_0^{(E)} \rangle. \label{eq:44-00}
\eea

The tree level diagrams are easy to compute. For example,
\be
\langle {\rm Tr} X_4 X_4 (\tau=0) \rangle|_{tree}=\int \frac{dp\, dq}{(2\pi)^2} 2\pi \delta(p+q)\frac{g^2_{PW}}{p^2 +4}N^2 = \frac{g^2_{PW}}{4}N^2.
\ee
Similarly,
\be
\langle {\rm Tr} X^{(E)} _0 X^{(E)}_0 (\tau=0)  \rangle|_{tree}=\frac{g^2_{PW}}{2}N^2.
\ee

\begin{figure}[ t]
\begin{center}
\includegraphics[width=11cm]{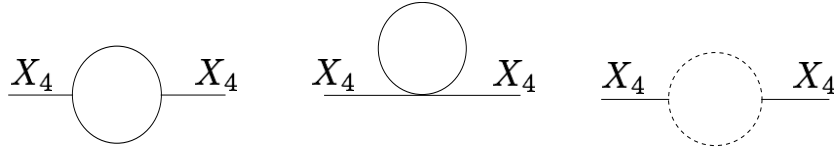}
\caption{The diagrams contributing to $\langle  {\rm Tr} \, X_4 X_4 \rangle$.  The dotted line represents fermion loop. The vertices of the left diagram come from the first term in \eqref{eq:int trivial}, which gives $\frac{1}{16}g^4_{PW} (N^3 -N)$. The vertices of the middle diagram come from the second and third terms in \eqref{eq:int trivial}, which give $-\frac{1}{32}g^4_{PW} (N^3 -N)$ and $-\frac{3}{16}g^4_{PW} (N^3 -N)$ . The vertices of the right diagram come from the fifth term in \eqref{eq:int trivial}. This diagram actually vanishes. }
\label{fig:X4X4}
\end{center}
\end{figure}

\begin{figure}[ t]
\begin{center}
\includegraphics[width=7cm]{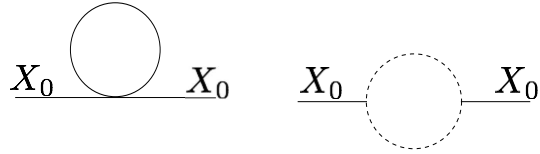}
\caption{The diagrams contributing to $\langle  {\rm Tr} \, X_0^{(E)}  X_0^{(E)} \rangle$. The vertices of the left diagram come from the third and the fourth terms in \eqref{eq:int trivial}, which give $-\frac{3}{8}g^4_{PW} (N^3 -N)$ and $-\frac{5}{4}g^4_{PW} (N^3 -N)$. The vertices of the right diagram come from the fifth term in \eqref{eq:int trivial}, which gives $\frac{5}{4}g^4_{PW} (N^3 -N)$. }
\label{fig:X0X0}
\end{center}
\end{figure}

 At the one-loop level, the diagrams  shown in Fig.~\ref{fig:X4X4} and Fig.~\ref{fig:X0X0} contribute. For example,  the vertices of the first diagram in Fig.~\ref{fig:X4X4} comes from the first terms in \eqref{eq:int trivial},
 \be
 \frac{i}{g^2_{PW}}\varepsilon_{a'b'c'} {\rm Tr}\,X_{a'}[X_{b'},X_{c'}]= \frac{6i}{g^2_{PW}}{\rm Tr}\,X_4 [X_2,X_3],
 \ee
 and thus, the diagram can be evaluated as
 \bea
 &&\frac{1}{2}\left( \frac{6i}{g^2_{PW}}\right)^2 \int \frac{dp_1 dp_2}{(2\pi)^2} \int \frac{dq_1 \cdots dq_6}{(2\pi)^6}(2\pi)\delta(q_1+q_2 +q_3) (2\pi) \delta(q_4+q_5+q_6) \nonumber\\
&&\ \ \ \times  \langle {\rm Tr}X_4(p_1) X_4 (p_2) \ {\rm Tr}X_4(q_1) [X_2 (q_2),X_3(q_3)]\  {\rm Tr}X_4 (q_4)[X_2(q_5),X_3(q_6)]  \rangle_{conn}\nonumber \\
  &=&\frac{1}{16}g^4_{PW} (N^3 -N) 
 \eea 
 The other  diagrams can be evaluated in a similar manner (for the contribution of each diagram, see the captions of Fig.~\ref{fig:X4X4} and Fig.~\ref{fig:X0X0}). The result is
  \bea
\langle  {\rm Tr} \, X_4 X_4 \rangle &=& \frac{g^2_{PW}}{4}N^2
+\left( \frac{1}{16}-\frac{1}{32}-\frac{3}{16} \right)g^4_{PW}(N^3-N)+ \mathcal{O}(g_{PW}^6)\\
 \langle   {\rm Tr} \, X_0^{(E)} X_0^{(E)} \rangle&=&\frac{g^2_{PW}}{2}N^2+ \left( -\frac{3}{8}-\frac{5}{4}+\frac{5}{4} \right)g^4_{PW}(N^3-N)+ \mathcal{O}(g_{PW}^6)
 \eea

Therefore, up to the one-loop order, we obtain
\bea
\langle  {\rm Tr} \, X_4 X_4 \rangle - \langle  {\rm Tr} \,  X_0^{(E)} X_0^{(E)} \rangle = -\frac{g^2_{PW} N^2}{4} + \frac{7}{32}(N^3 -N )g^4_{PW} + \mathcal{O}(g_{PW}^6). \label{eq:pert pwmm}
\eea

\subsection*{One-loop calculation in matrix integral}\label{sec:1loopmatrix}
We can apply the localization method to  compute the observable \eqref{eq:PWMMobs} around the trivial background  in the PWMM. The saddle point configuration corresponding to the trivial background is given by $\hat{X}_{a'}=0$
in (\ref{QV=0}).
In this case, since $\hat{X}^{(E)}_0(\tau=0)=M$, 
\eqref{eq:PWMMobs} is reduced to the following matrix integral,
\be
\langle {\rm Tr}  \,  (i M) ^2  \rangle = - \int \left( \prod_{i=1}^N dm_i\right) \sum_{i=1}^N (m_i)^2 \  Z_{1-loop}^{(trivial)}\ \exp\left(-\frac{2}{g^2_{PW}} \sum_{i} (m_i)^2\right) ,\label{eq:EVobs}
\ee
where $  Z_{1-loop}^{(trivial)} $ is the determinant factor \eqref{1loopdet} for the trivial background;
\be
Z_{1-loop}^{(trivial)} = \prod_{i<j} \frac{(4+(m_i - m_j)^2)(m_i - m_j)^2}{(1+(m_i - m_j)^2 )^2}. \label{eq:det trivial}
\ee

In order to perform  a perturbative calculation, we express the above eigenvalue integral into a covariant form. Firstly,  the factor $\prod_{i<j}(m_i- m_j)^2$ in \eqref{eq:det trivial} gives the correct measure of the hermitian matrix integral,
\be
\int \left( \prod_{i=1}^N dm_i\right) \prod_{i<j}(m_i- m_j)^2 = \int d M.
\ee
For the other part of  \eqref{eq:det trivial}, we exponentiate it as
\bea
\prod_{i<j}\frac{4+(m_i - m_j)^2}{(1+(m_i - m_j)^2)^2}= \exp\biggl[\sum_{i<j} \log(4+(m_i - m_j)^2 )-2 \sum_{i<j}\log(1+ (m_i-m_j)^2)\biggr ].
\eea
The first term in the exponent can be written in terms of the matrix $M$ (up to the irrelevant constant $ N(N-1)\log 4$) as
\bea
\sum_{i<j} \log[1+\frac{1}{4}(m_i - m_j)^2] 
= \frac{1}{2}\sum_{i,j}\sum_{n=1}^{\infty}\frac{(-1)^{n+1}}{n \cdot 4^n}\sum_{r=0}^{2n} \begin{pmatrix}
2n\\
r\\
\end{pmatrix} (m_i)^{2n-r}(-m_j)^{r}\nonumber\\
= \frac{1}{2}\sum_{i,j}\sum_{n=1}^{\infty}\frac{(-1)^{n+1}}{n \cdot 4^n}\sum_{r=0}^{2n} \begin{pmatrix}
2n\\
r\\
\end{pmatrix}(-1)^r {\rm Tr} M^{2n-r} {\rm Tr} M^r.
\eea
Similarly, the second term in the exponent can also be written as
\bea
-2 \sum_{i<j}\log(1+ (m_i-m_j)^2)=\sum_{n=1}^{\infty}\frac{(-1)^{n}}{n }\sum_{r=0}^{2n} \begin{pmatrix}
2n\\
r\\
\end{pmatrix}(-1)^r {\rm Tr} M^{2n-r} {\rm Tr} M^r.
\eea
Thus,  we can express the eigenvalue integral as the following matrix model,
\bea
\int dM\ e^{S[M]},
\eea
where the action $S[M]$ is 
\bea
S=-\frac{2}{g^2_{PW}} {\rm Tr} M^2 + \sum_{n=1}^{\infty} \sum_{r=0}^{2n}C_{n,r} {\rm Tr} M^{2n-r} {\rm Tr} M^r. \label{eq:trivialMM}
\eea
Here, the coefficients $C_{n,r}$ are given by
\bea
C_{n,r}= \frac{(-1)^{n+r}}{n}\begin{pmatrix}
2n\\
r\\
\end{pmatrix}\left(1-\frac{1}{2\cdot 4^n}\right).
\eea
The propagator is given by
\bea
\langle M_{ij} M_{kl} \rangle = \frac{g^2_{PW}}{4}\delta_{il}\delta_{jk}.
\eea
 
We compute the observable \eqref{eq:EVobs} up to the one-loop order.  The tree level contribution is given by $-\frac{g^2_{PW} N^2}{4}$. Note that at the 
one-loop level, the relevant interactions  in \eqref{eq:trivialMM} are 
only the terms with $n=1$,
\bea
(C_{1,0}+C_{1,2})\, {\rm Tr} \, {\bf 1}_{N\times N} \, {\rm Tr} M^2 + C_{1,1}{\rm Tr}M \,{\rm Tr}M= \frac{7}{4}(-N\,{\rm Tr} M^2 +{\rm Tr}M \,{\rm Tr}M).
\eea
Then we can easily find
\bea
\langle {\rm Tr}  \,  (i M) ^2  \rangle =-\frac{g^2_{PW} N^2}{4} + \frac{7}{32}(N^3 -N )g^4_{PW} + \mathcal{O}(g_{PW}^6).
\eea
This agrees completely with the result obtained from the original PWMM, \eqref{eq:pert pwmm}.

\end{document}